\documentclass[10pt,twocolumn]{article}
\usepackage{simpleConference}
\usepackage{times}
\usepackage{float}
\usepackage{graphicx}
\usepackage{caption}
\usepackage{subcaption}
\usepackage{amsmath,amssymb,amsfonts,amsthm}
\usepackage{url,hyperref}
\usepackage{xcolor}
\usepackage{tabularx}
\usepackage{fancyhdr}
\usepackage{braket}
\usepackage[linesnumbered,ruled,vlined]{algorithm2e}

\begin{document}

\title{ScaleQC: A Scalable Framework for Hybrid Computation on Quantum and Classical Processors\thanks{Code available at: https://doi.org/10.5281/zenodo.6421908}}

\author{Wei Tang\\
Department of Computer Science\\
Princeton University\\
\texttt{weit@princeton.edu}
\and
Margaret Martonosi\\
Department of Computer Science\\
Princeton University\\
\texttt{mrm@princeton.edu}
}

\maketitle
\thispagestyle{empty}

\begin{abstract}
  Quantum processing unit (QPU) has to satisfy highly demanding quantity and quality requirements on its qubits to produce accurate results for problems at useful scales.
  Furthermore, classical simulations of quantum circuits generally do not scale.
  Instead, quantum circuit cutting techniques cut and distribute a large quantum circuit into multiple smaller subcircuits feasible for less powerful QPUs.
  However, the classical post-processing incurred from the cutting introduces runtime and memory bottlenecks.
  Our tool, called ScaleQC, addresses the bottlenecks by developing novel algorithmic techniques including
  (1) a quantum states merging framework that quickly locates the solution states of large quantum circuits;
  (2) an automatic solver that cuts complex quantum circuits to fit on less powerful QPUs;
  and (3) a tensor network based post-processing that minimizes the classical overhead.
  Our experiments demonstrate both QPU requirement advantages over the purely quantum platforms,
  and runtime advantages over the purely classical platforms
  for benchmarks up to $1000$ qubits.
\end{abstract}

\section{Introduction}
Quantum Computing (QC) has been proposed as a promising counterpart to classical computing, but QPUs face demanding hardware requirements to operate at useful scales.
Many quantum algorithms offer runtime advantages over the best known classical algorithms, such as unstructured database search~\cite{grover1996fast} and integer factorization~\cite{shor1999polynomial}.
To solve a problem, researchers develop a quantum algorithm that is represented as a $n$ qubit quantum circuit.
A QPU then executes the circuit and samples its probability output to find the ``solution'' states,
indicated by their much higher quantum state amplitudes than the non-solution states.
In order to even run the circuit at all, the QPU must have at least $n$ qubits.
Furthermore, the QPU's qubits must be accurate and robust enough to support the quantum workload without accumulating too much noise to produce quality solutions.
However, these two requirements put a heavy toll on the hardware.
As an example, the famous Shor's integer factorization algorithm is expected to take millions of physical qubits in order to construct enough number of high quality logical qubits to run problems at practical scales~\cite{suchara2013comparing}.

Instead, the recent introduction of the quantum circuit cutting theory~\cite{peng2020simulating} and its early demonstrations~\cite{tang2021cutqc, tang2022cutting}
hint at the potential to combine multiple less capable QPUs and classical post-processing into a hybrid architecture.
Circuit cutting techniques divide a large quantum circuit into several smaller subcircuits,
which can be executed in parallel on multiple QPUs with lower qubit quantity and quality requirements.
Classical post-processing is then used to reconstruct the output from these small subcircuit outputs.
This hybrid architecture is analogues to the classical parallel computing where a large workload is distributed among many computing nodes,
which afford to be much less powerful,
but at the expense of some communication cost to reconstruct the results.

Dividing a large quantum circuit into many smaller subcircuits introduces obvious advantages over relying on either quantum or classical platforms alone.
The most important advantage comes from that the subcircuits usually have fewer quantum gates, potentially also have fewer qubits.
This puts a much lower qubit quality requirement for QPUs to produce accurate results, as the shallower and less complicated subcircuits incur much less crosstalk~\cite{mundada2019suppression}, decoherence~\cite{klimov2018fluctuations}, gate errors~\cite{arute2019quantum} and control difficulties~\cite{abdelhafez2020universal}.
In addition, having fewer qubits directly reduces the strict size limit on QPUs.
Specifically, \cite{tang2021cutqc} demonstrates that certain algorithms can be run with fewer than half the number of qubits.
Furthermore, since QPUs operate much faster than the classical simulations of quantum circuits, such hybrid architecture offers significant runtime speedup.
Overall, quantum circuit cutting techniques combine less powerful QPUs and classical computing to expand the computational reach for both.

Despite the obvious advantages, two key challenges remain to make quantum circuit cutting practical.
First, the classical post-processing overhead grows quickly with the number of cuts and bottlenecks the end-to-end workflow, preventing its application to hard-to-cut circuits.
Second, the memory overhead to store and compute the state space of quantum circuits still doubles with every additional qubit, preventing its application to large circuits.

This paper overcomes the fundamental runtime and memory scalability difficulties of quantum circuit cutting and makes the hybrid architecture practical.
Via an iterative search framework called states merging,
ScaleQC offers the ability to locate solution states for arbitrarily large quantum circuits,
hence bypassing the classical memory limitations.
Via an automatic cut solver,
ScaleQC finds high quality cuts suitable for post-processing and generalizes to different benchmarks.
Via the compute graph contraction algorithm,
ScaleQC reduces the runtime overhead to enable the applications to complicated circuits.

To evaluate the performance,
we ran several realistic quantum benchmark algorithms such as QAOA~\cite{saleem2020approaches},
AQFT~\cite{barenco1996approximate},
Maximum Independent Set~\cite{saleem2020approaches},
Supremacy~\cite{arute2019quantum} and BV~\cite{bernstein1997quantum}.
We demonstrated running quantum circuits of up to $1000$ qubits.
Our contributions include the following:
\begin{enumerate}
    \item \textbf{Reduce Memory}: Developed a states merging method to efficiently locate arbitrary solution states for large quantum circuits, thus overcoming the classical memory scalability obstacles.
    \item \textbf{Reduce Runtime}: Developed a Mixed Integer Programming (MIP) solver that automatically finds cuts for large quantum circuits to allow easier classical post-processing,
    while constrained by user-specified available QPU resources.
    \item \textbf{Reduce Runtime}: Developed a compute graph contraction algorithm to significantly reduce the classical runtime overhead.
\end{enumerate}
\section{Background}
This section introduces the quantum circuit cutting theory,
its use cases and identifies its key challenges.

\subsection{Circuit Cutting Theory}
\begin{figure}[t]
    \centering
    \begin{subfigure}{0.4\textwidth}
        \centering
        \includegraphics[width=\textwidth]{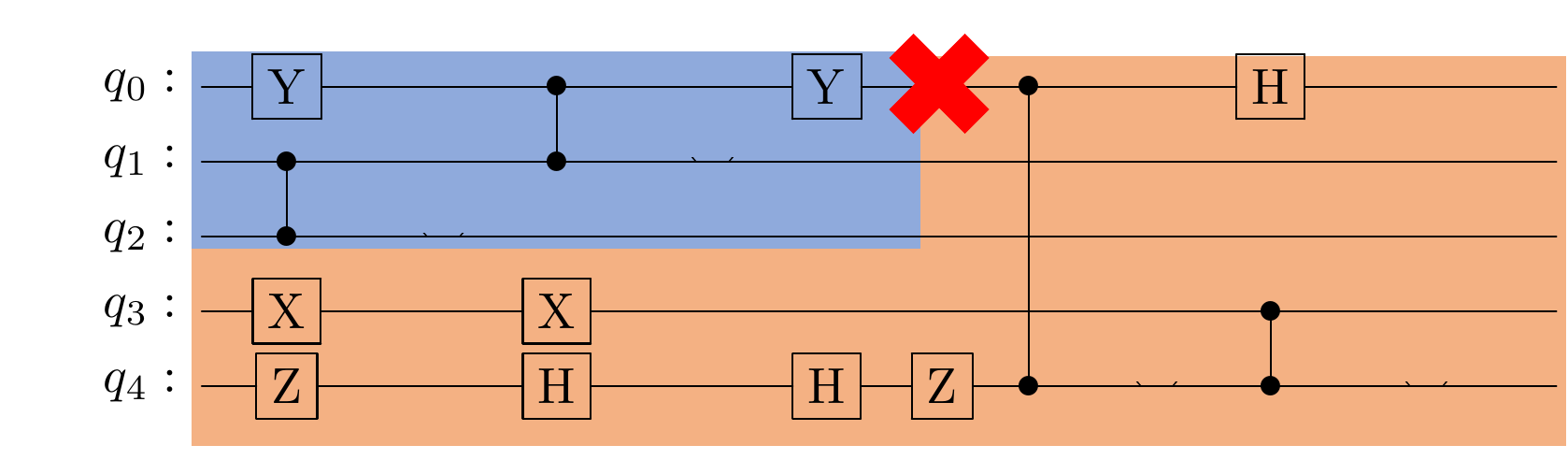}
        \caption{The red cross indicates the cutting point.
        Subcircuit $1$ is shaded blue and subcircuit $2$ is shaded orange.}
        \label{fig:simple_circuit}
    \end{subfigure}
    \begin{subfigure}{0.4\textwidth}
        \centering
        \includegraphics[width=\textwidth]{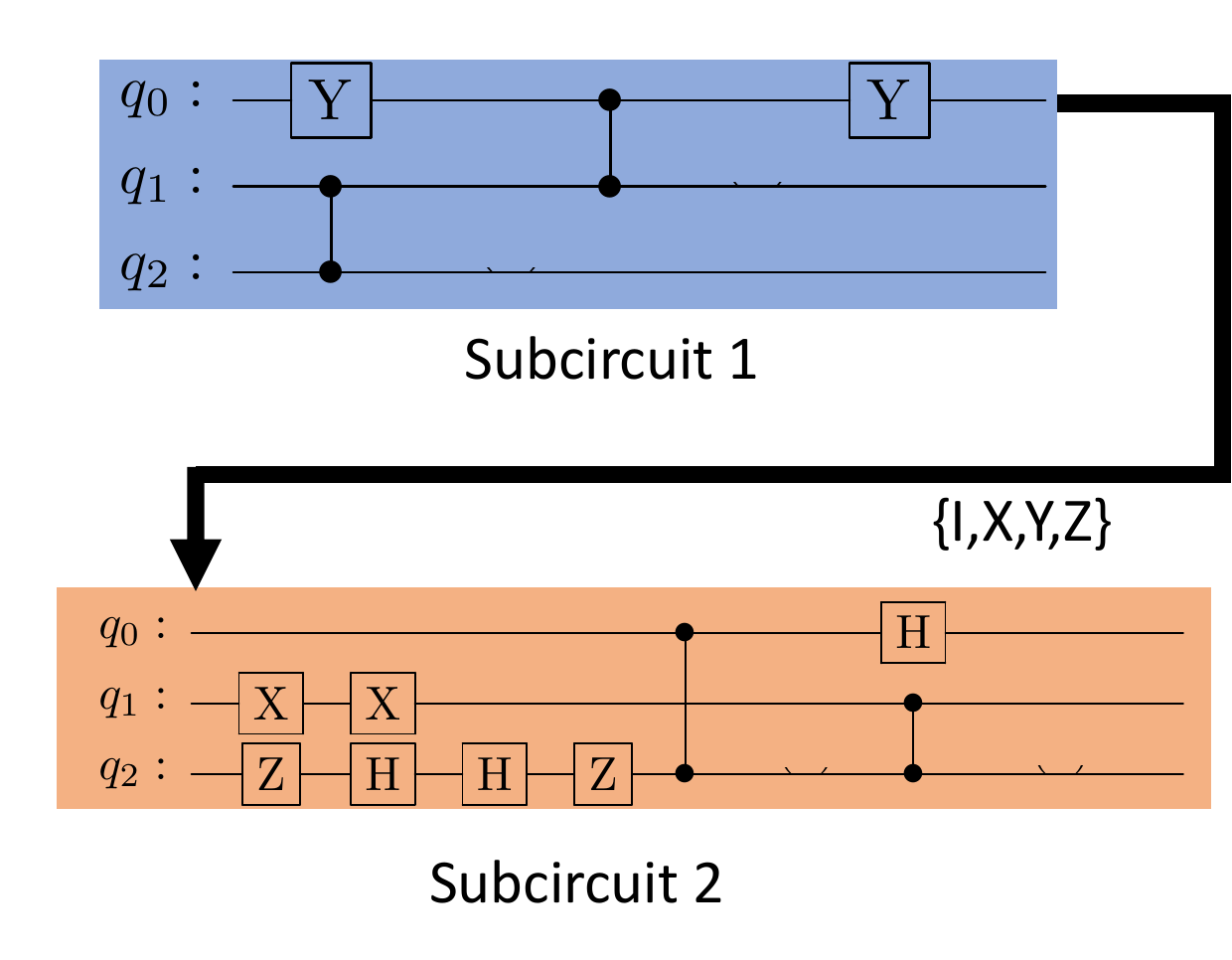}
        \caption{The black arrow between the subcircuits shows the ``path'' undertaken by the qubit line being cut.
        The one cut needs to permutate each of the $\{I,X,Y,Z\}$ bases.
        Subcircuit $1$ needs to be measured in each basis and 
        subcircuit $2$ needs to be initialized in each basis.
        The two subcircuits can now be executed independently on $3$-qubit QPUs.}
        \label{fig:simple_cutting}
    \end{subfigure}
    \caption{Example of cutting a $5$-qubit quantum circuit with $1$ cut that divides it into two smaller subcircuits.}
    \label{fig:simple_cutting_example}
\end{figure}
While we refer the readers to~\cite{peng2020simulating} for a detailed derivation and proof of the physics theory,
we provide an intuitive understanding of the cutting process in order to identify its key challenges.

Circuit cutting cuts between quantum gates and decomposes the qubit states into their Pauli bases.
With a proper selection of the cutting points, a large quantum circuit can be divided into smaller isolated subcircuits.
Figure~\ref{fig:simple_circuit} shows an example $5$-qubit quantum circuit.
Without cutting, this circuit requires a QPU with at least $5$ good enough qubits to execute all the quantum gates before too many errors accumulate.
Circuit cutting makes $1$ cut and divides this quantum circuit into two smaller subcircuits,
each with both fewer qubits and fewer gates.
Figure~\ref{fig:simple_cutting} shows that each subcircuit needs to be measured and initialized in different bases according to the permutations of all the cuts.
Now multiple less powerful 3-qubit QPUs can run these independent subcircuits in parallel.
The quantum interactions among the subcircuits are substituted by classical post-processing,
which are analogues to the communication cost paid in classical parallel computing.

In general, a $n$ qubit quantum circuit undergoes $K$ cuts to divide into $n_C$ completely separated subcircuits $C=\left\{C_1,\ldots,C_{n_C}\right\}$.
A complete reconstruction of the quantum interactions requires each cut to permute each of the $\{I,X,Y,Z\}$ bases, for a total of $4^K$ assignments.
Depending on the basis assigned to the cuts connecting each subcircuit, it is initialized and measured slightly differently to produce a distinct entry.
We use $p_i(k)$ to represent the output of subcircuit $i$ in the $k$th cuts assignment, where $i\in\{1,\ldots,n_C\}$ and $k\in\{1,\ldots,4^K\}$.
The physics theory~\cite{peng2020simulating} dictates that the output of the original circuit is given by
\begin{equation}
    P=\sum_{k=1}^{4^K}\otimes_{i=1}^{n_C}p_i(k)\in\mathbb{R}^{2^n}\label{eq:CutQC}
\end{equation}
where $\otimes$ is the outer product between two subcircuit output vectors.
Note that each subcircuit only has a different entry when at least one cut connecting it in the compute graph changes basis.
This means that many of the $p_i(k)$ repeat in the various summation terms.
\subsection{Circuit Cutting Use Cases}
Several theory proposals~\cite{bravyi2016trading, peng2020simulating} first introduce the possibility of quantum circuit cutting.
Many small-scale demonstrations exist for chemical molecule simulations~\cite{eddins2021doubling} and variational quantum solvers~\cite{yuan2021quantum}.
In addition, several quantum computing industry leaders such as IBM and Xanadu~\cite{Xanadu} are actively developing quantum circuit cutting products.
These efforts either apply the hybrid architecture to real-world problems,
or aim at incorporating it into the existing quantum hardware and software stack.
However, none of the current efforts addresses the scalability obstacles.
\subsection{Circuit Cutting Challenges}\label{sec:challenges}
The key challenges of applying quantum circuit cutting at useful scales lie with the classical post-processing.
Equation~\ref{eq:CutQC} clearly shows the two challenges as the following:

\begin{enumerate}
    \item The length of the full output of a $n$-qubit circuit is $2^n$.
    Large quantum circuits quickly bottleneck the classical memory as well as the runtime.
    \item The classical post-processing scales exponentially with the number of cuts required $K$ and hence bottlenecks the runtime.
\end{enumerate}

This paper develops the states merging technique to bypass the memory challenge altogether while reducing the classical overhead.
In addition, this paper proposes compute graph contraction to significantly reduce the runtime overhead.
\section{Framework Overview}
\begin{algorithm}[t]
    \DontPrintSemicolon
    \SetAlgoLined
    \caption{ScaleQC Framework}\label{alg:framework}
    \KwIn{Quantum circuit $C$.
    Max load factor $\alpha$.
    Max number of probability bins allowed per recursion $M$.
    Max number of recursions $R$.}
    Initialize an empty list $L$\;
    Recursion counter $r\gets0$\;
    Find cuts for $C$ to satisfy the max load $\alpha$ and produce the compute graph $G$. \tcp*[l]{Section~\ref{sec:locate_cuts}}
    $n_i\gets$ number of data qubits in subcircuit $i\in\{1,\ldots,n_C\}$\;
    \While{$r<R$}{
        \uIf{$r>0$ \textbf{and} $bin$ is None}{
            BREAK\tcp*[l]{All solution(s) are found}
        }
        Compute the bin assignments for each subcircuit \tcp*[l]{Algorithm~\ref{alg:states_assignment}[$r$,$bin$,$M$,$n_{1\ldots n_C}$], Section~\ref{sec:states_merging}}
        \textbf{QPU} : Run the subcircuits to produce the merged probability bins for the subcircuit $\{p_{i,k}\}$\;
        Contract the compute graph $G$ to reconstruct the probability sums for the bins \tcp*[l]{Section~\ref{sec:compute_graph_contraction}}
        Append the $R$ largest bins not yet fully expanded to $L$\;
        Sort and truncate $L$ to keep the largest $R$ bins\;
        Pop $bin$ from $L$\;
        $r\gets r+1$\;
    }
\end{algorithm}

Algorithm~\ref{alg:framework} outlines the overall framework of ScaleQC.
The input to the framework includes the quantum circuit itself
and a list of hyperparameters:
\begin{enumerate}
    \item Max QPU load factor $\alpha$: The maximum fraction of the two-qubit gates each QPU can handle.
    For example, $\alpha=0.5$ for a quantum circuit with $100$ two-qubit gates means ScaleQC needs to cut into subcircuits with at most $50$ two-qubit gates.
    \item Max per-recursion bins $M$: The maximum number of probability bins each states merging recursion computes.
    \item Max recursions $R$: maximum number of states merging recursions to run.
\end{enumerate}

The framework is mainly built around an iterative procedure called the states merging to locate the solution states for the input quantum circuit
(the $while$ loop in ALgorithm~\ref{alg:framework}).
Section~\ref{sec:states_merging} discusses the iterative procedure in detail.
On a high level, ScaleQC recursively evaluates the subcircuits with QPUs and searches the solution states with classical computing until all solution states are found,
or the user-defined max recursion depth $R$ is reached first.
States merging narrows down to a range of possible states containing the solutions when the max recursion depth is reached first.

While one can easily use a lot of cuts to separate a large circuit to satisfy the size constraints,
the scalability and efficiency of the post-processing largely rely on the strategic choice of the cutting points.
Section~\ref{sec:locate_cuts} discusses the MIP solver used to find high quality cuts for any input quantum circuits.
Applying the cuts to the input quantum circuit then produces a compute graph abstraction for post-processing.

Quantum and classical platforms then use the compute graph generated from the cuts to reconstruct the original quantum circuit output.
Section~\ref{sec:compute_graph_contraction} establishes the classical post-processing of circuit cutting as tensor network contractions,
and hence introduces a compute graph contraction method running on GPUs for an efficient classical post-processing.
\section{States Merging}\label{sec:states_merging}
\begin{algorithm}[t]
    \DontPrintSemicolon
    \SetAlgoLined
    \caption{States Assignment\;
    \textbf{Goal}: Reduce memory overhead.}\label{alg:states_assignment}
    \KwIn{
        $r$, $bin$, $M$, $n_{1\ldots n_C}$\tcp*[l]{Refer Algorithm~\ref{alg:framework}}}
    \uIf{$r=0$}{\label{curr_num_bins_start}
        Initialize each quantum state in its own bin for $l_i\gets2^{n_i}$ bins in subcircuit $i$, $i\in\{1,\ldots,n_C\}$\;
    }
    \uElse{
        Locate the subcircuit states contained in $bin$, $l_i\gets\#states_i$\;
    }\label{curr_num_bins_end}
    Copy $l'_i\gets l_i$\;
    Initialize the total number of bins required $l\gets\prod_i{l'_i}$\;\label{target_num_bins_start}
    \While{$l>M$}{
        Pick the subcircuit $i$ with the largest $l'_i$\;
        $l'_i\gets \lceil l'_i/2\rceil$\;
        $l\gets\prod_i{l'_i}$\;
    }\label{target_num_bins_end}
    \For{$i\in\{1,\ldots,n_C\}$}{\label{assign_states_begin}
        Initialize $l'_i$ empty bins\;
        \For{$j=1,\ldots,l_i$}{
            $k\gets j\%l'_i$\;
            Append the $j$th state of subcircuit $i$ to bin $k$ of subcircuit $i$\;
        }
    }\label{assign_states_end}
    \Return{Updated subcircuit bin assignment}
\end{algorithm}

\begin{figure}[t]
    \centering
    \includegraphics[width=.6\linewidth]{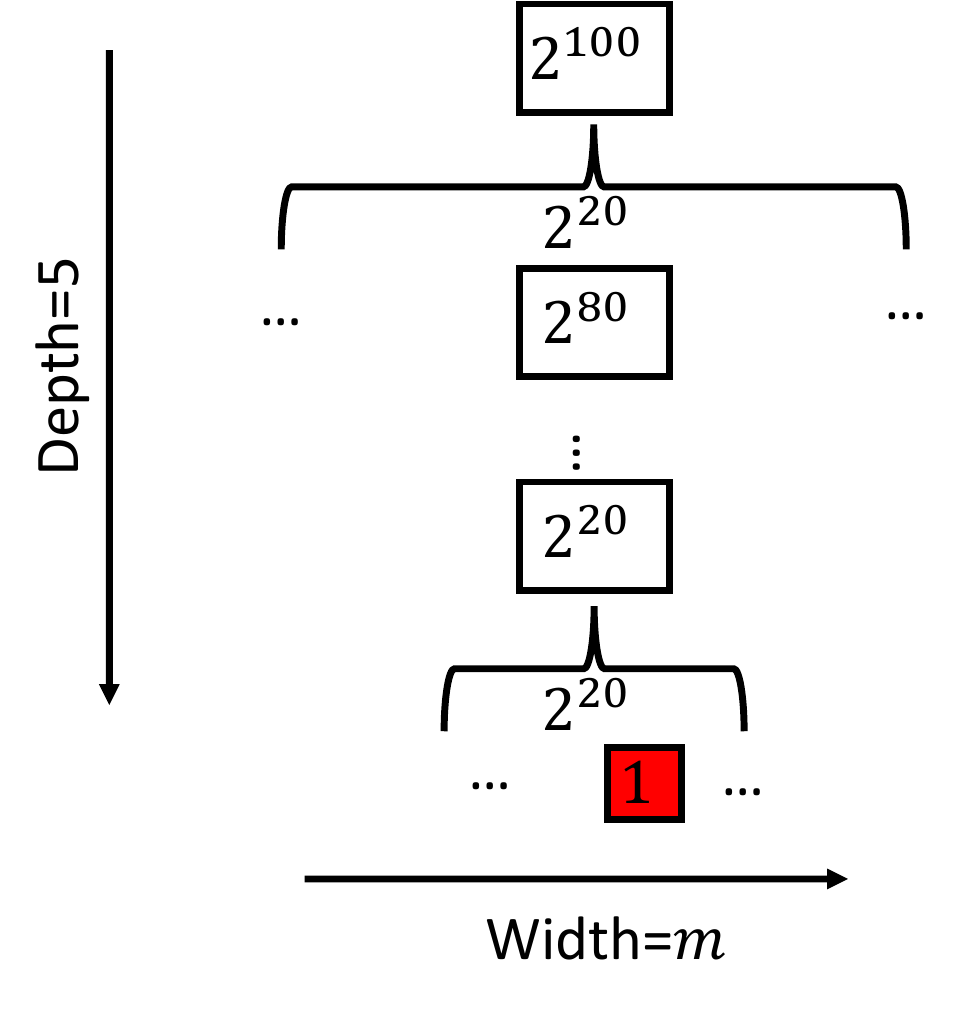}
    \caption{States merging locates solution states for a $100$-qubit circuit,
    with $M=2^{20}$.
    Each box represents a bin.
    The boxed numbers represent the number of quantum states in the bin.
    Each bracket represents a recursion.
    The numbers on the brackets indicate the number of bins for the recursion.
    The red box represents a solution state found.}
    \label{fig:states_merging}
\end{figure}

Quantum algorithms can be loosely characterized as two groups based on the type of their output.
The first type produces concentrated outputs, where a few solution states have much higher probability amplitudes than the others.
The solution states should only occupy a very small part of the entire state space,
which is the very reason that it is useful for the quantum algorithms to find them efficiently.
This paper mainly focuses on demonstrating circuits of this type.

The second type of algorithms produces a certain probability distribution,
where the probability amplitudes are much more distributed among the states.
We argue that large size circuits of this type are not practical,
since their probability distributions most likely require exponentially many shots to converge even with a large and reliable QPU.
In addition, it is impossible to store or analyze the entire state space.
As a result, they are usually limited at medium sizes.
ScaleQC is able to produce the full state output for this type of circuits.

One major bottleneck of quantum circuit cutting is that the length of the probability output of a $n$-qubit quantum circuit is $2^n$.
Hence, full state reconstruction for large circuits quickly increases both the memory requirements and runtime.
As an alternative, states merging offers a scalable way to only locate the solution states for large circuits.

The states merging method makes use of the fact that the individual probabilities of the non-solution states are not of interest.
Hence, instead of keeping track of the individual probability for each quantum state,
the states merging method merges multiple states into one bin and keeps track of their sum of probabilities.
The bins containing the solution states hence have much higher probabilities than the other bins and can be easily identified.
The \textit{while} loop in the ScaleQC framework in Algorithm~\ref{alg:framework} shows the complete process.

Each recursion of the procedure first assigns the quantum states still under analysis into various bins.
Algorithm~\ref{alg:states_assignment} shows the assignment process.
Lines~\ref{curr_num_bins_start}-\ref{curr_num_bins_end} first calculate the current number of bins for each subcircuit.
The first recursion starts with each quantum state in its own separate bin,
hence each subcircuit has $l_i=2^{n_i}$ bins,
where $n_i$ is the number of data qubits in the subcircuit $i$.
Subsequent recursions seek to expand a particular selected $bin$,
hence each subcircuit has the number of subcircuit states belonging to the selected $bin$.

Next, lines~\ref{target_num_bins_start}-\ref{target_num_bins_end} calculate the target number of bins to use for the next recursion.
We shrink the subcircuit with the most number of bins by $2$.
The shrinking steps repeat until the total number of bins required satisfies the given memory limit $M$.

Finally, lines~\ref{assign_states_begin}-\ref{assign_states_end} evenly assign the quantum states under analysis into the number of bins calculated for the next recursion.

The max number of bins $M$ enables a key trade-off for the states merging procedure.
The states merging procedure is equivalent to a full state reconstruction if the max number of bins allowed $M$ is larger than the size of the full state $2^n$.
In this case, the states merging procedure simply locates all the solution states in one recursion.
Any $M$ smaller than $2^n$ essentially provides a trade-off of the size of the vectors to compute during each iteration versus the number of iterations required to locate all the solution states.

For the extreme case where the non-solution states have $0$ probability,
the states merging procedure is guaranteed to
find all the solutions in at most $mn/\log_2{M}$ recursions for a $n$-qubit circuit.
Where $m$ is the number of solution states.
Figure~\ref{fig:states_merging} visualizes an example for a $100$-qubit circuit,
assuming $M=2^{20}$ bins,
which is negligible as compared to the $2^{100}$ full states.
Each recursion expands a bin for finer details of the quantum states contained.
A max depth of $\frac{100}{20}=5$ produces leaf bins with just $1$ state.
A solution state is found when a leaf bin with just $1$ state and high probbaility is computed.
The entire states merging process hence creates a search tree with
a depth of $n/\log_2{M}$ and width of $m$,
representing total $mn/\log_2{M}$ recursions.

The procedure finds solutions more quickly if multiple solution states happen to be assigned into the same bin,
which reduces the width of the search tree at some depth.
Having larger $M$ shortens the search tree but has diminishing returns,
as doubling $M$ only increases the denominator by $1$.
Furthermore, larger $M$ means slower post-processing for each recursion.
As a result, it is usually favored to use smaller $M$ even if larger $M$ fits in the memory.

In addition, larger quantum circuits (larger $n$) and more solution states (larger $m$)
only increase the number of states merging recursions linearly.
The method hence easily scales to large quantum circuits with more solution states.

\textbf{Comparison with classical arbitrary state simulation methods:}
The state-of-the-art simulation work~\cite{liu2021closing} only simulates an arbitrary subset of quantum states for large circuits
to avoid the exponential number of quantum states.
Our method easily achieves the same functionality
if we just select an arbitrary subset of quantum states to run for a single recursion without merging into bins.
However, such arbitrary selection methods will have virtually $0$ chance to include any solution states in the sampled subset.
To find any solution states, such methods will need to run almost $2^n$ times and thus do not scale.
Instead, the states merging method strictly outperforms the existing methods via its iterative searches.

\cite{tang2021cutqc} proposed a Dynamic Definition search method that is similar in spirit with states merging.
The method iteratively fixes the state of the individual qubits to search for the solution states.
However, this method forces quantum states without common qubit states (such as $\ket{01}$ and $\ket{10}$) to be in different bins,
thus can only be located in separate recursions.
States merging is hence more efficient as it allows any solution states to be located simultaneously.
\section{Locate Cutting Points}\label{sec:locate_cuts}
\begin{figure*}[!t]
    \centering
    \begin{subfigure}{0.5\textwidth}
        \centering
        \includegraphics[width=\textwidth]{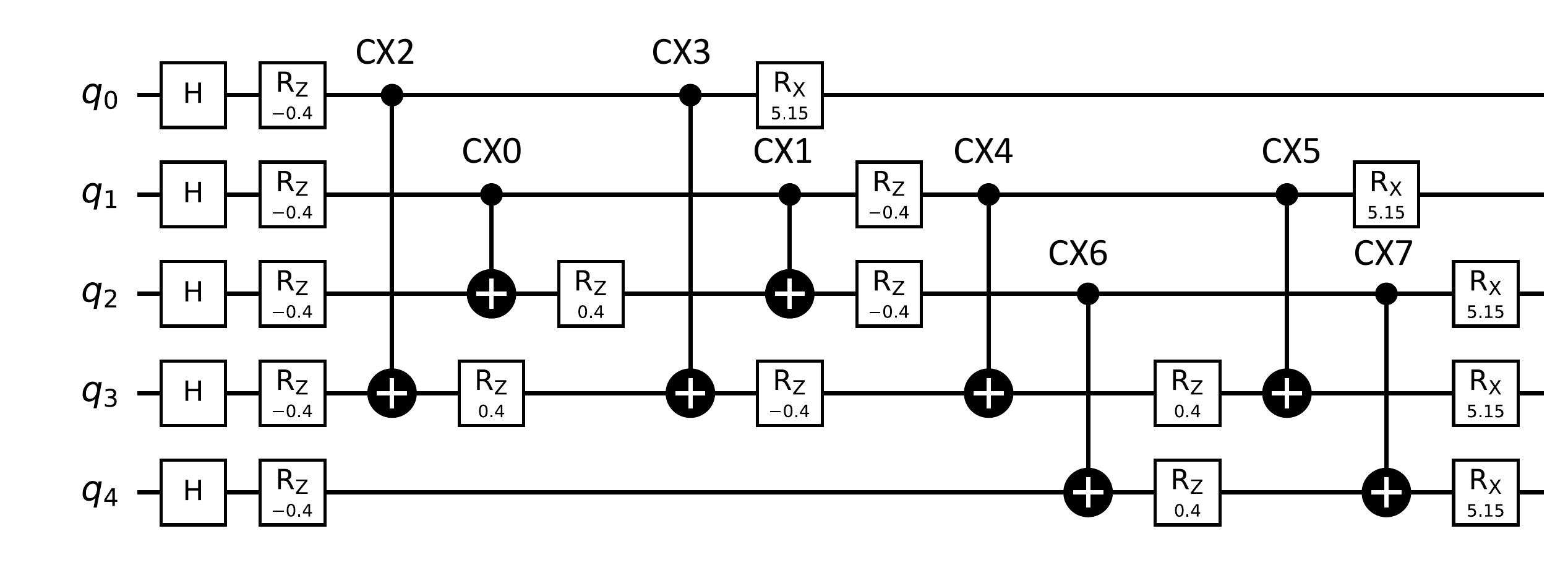}
        \caption{Quantum circuit.
        Each horizontal line is an input qubit.
        Time goes from left to right.
        The indices on the two-qubit $CX$ gates map across the subfigures.}
        \label{fig:circuit}
    \end{subfigure}
    \begin{subfigure}{0.4\textwidth}
        \centering
        \includegraphics[width=\textwidth]{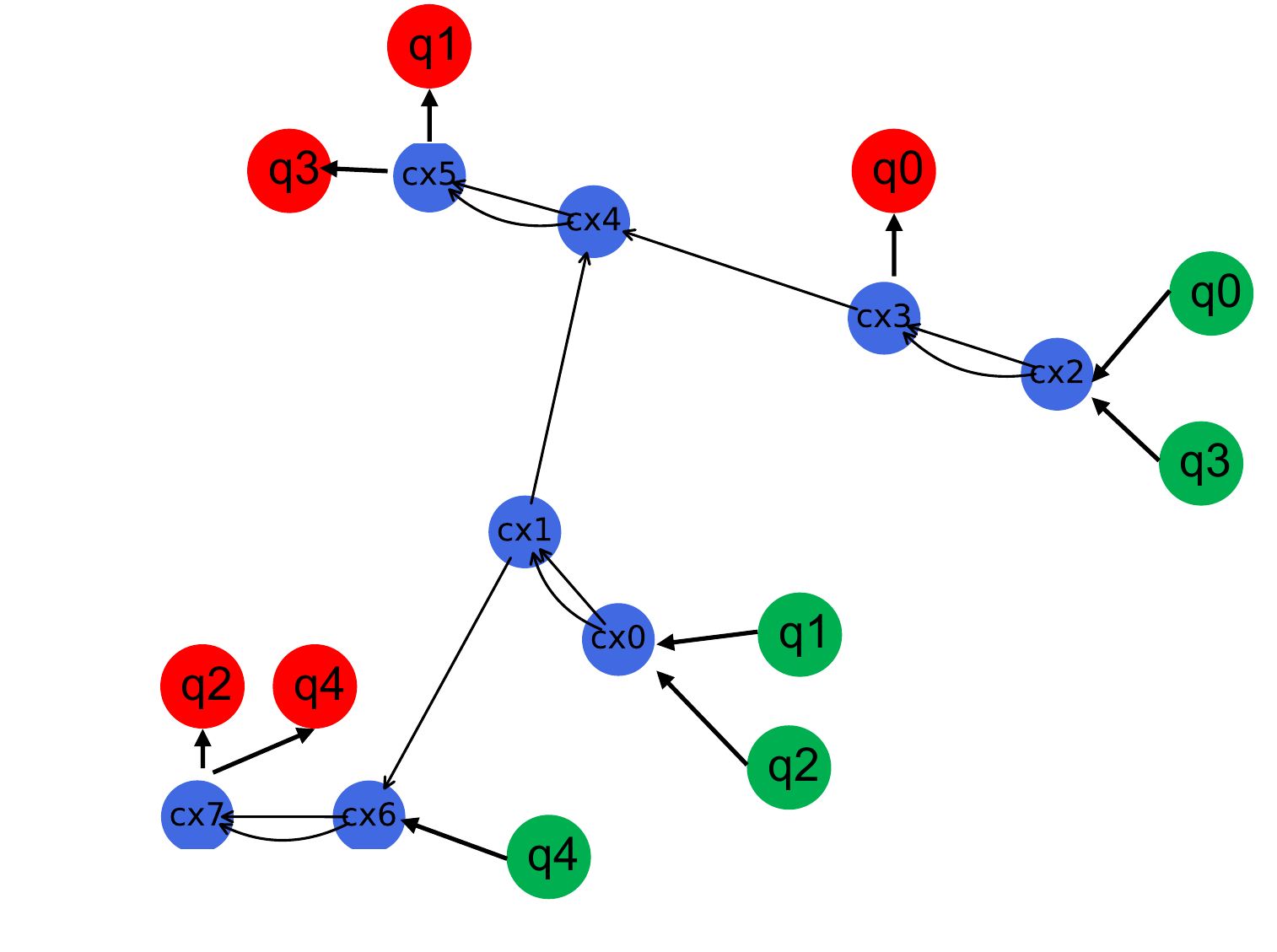}
        \caption{DAG representation of~\ref{fig:circuit} without the single-qubit gates.
        The green vertices indicate the input qubits.
        The red vertices indicate the output qubits.
        The blue vertices indicate the two-qubit quantum gates.
        The directed edges indicate the time flow of the qubit lines.
        Qubits in the initial states (green) go through the quantum gates (blue) and measured as output (red).}
        \label{fig:dag}
    \end{subfigure}
    \newline
    \begin{subfigure}{0.4\textwidth}
        \centering
        \includegraphics[width=\textwidth]{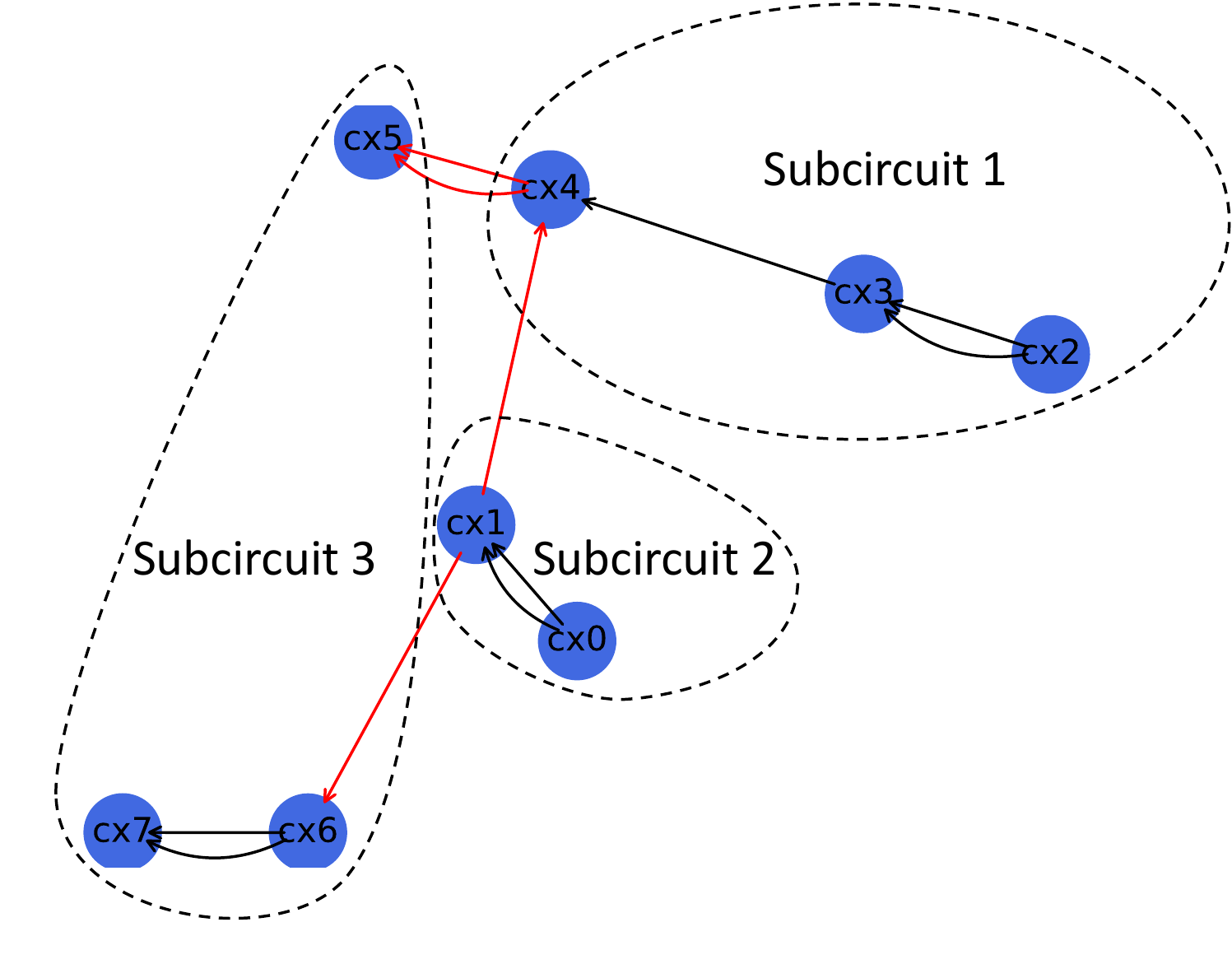}
        \caption{We only consider the blue quantum gates from~\ref{fig:dag} for the DAG partition problem.
        The red edges indicate the cuts found by our solver, subject to $\alpha=0.4$ max load constraint, i.e. $3$ two-qubit gates.}
        \label{fig:cut_dag}
    \end{subfigure}
    \begin{subfigure}{0.4\textwidth}
        \centering
        \includegraphics[width=\textwidth]{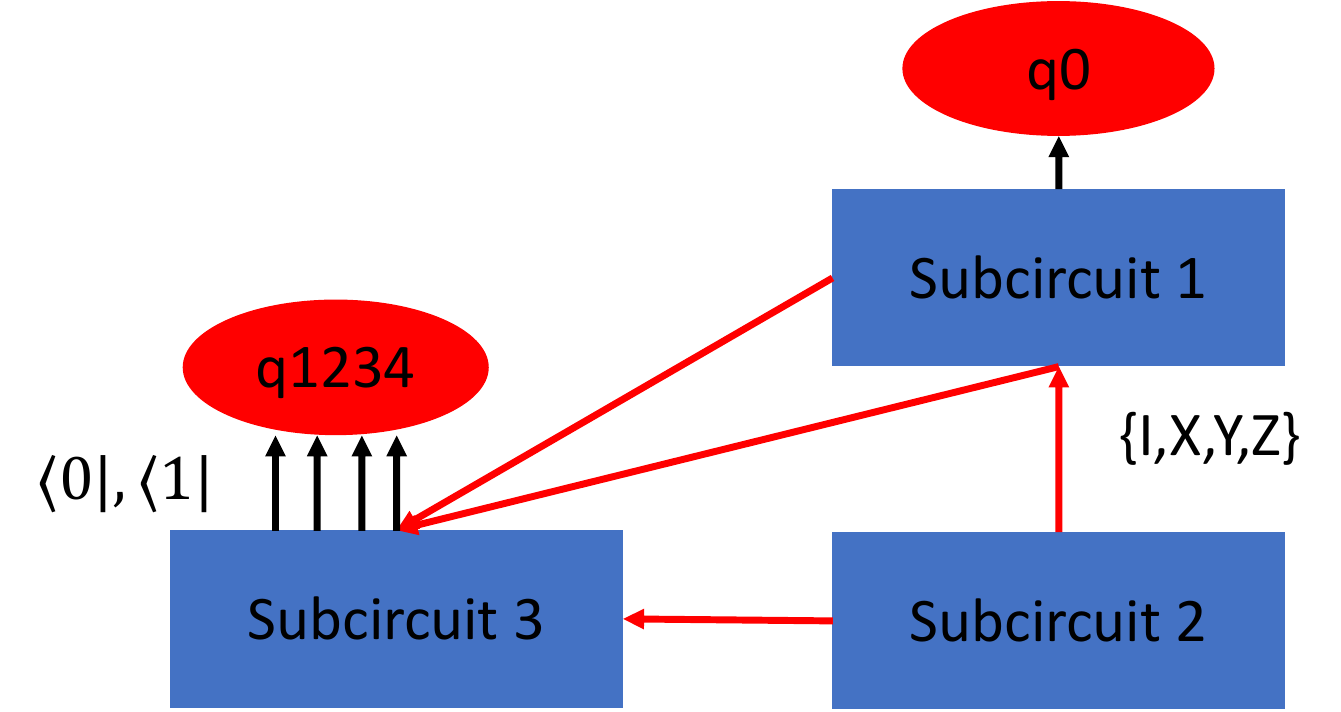}
        \caption{Compute graph abstraction.
        The cut (red) edges among the subcircuits are the indices to be contracted.
        The output qubit (black) edges are the output indices.}
        \label{fig:compute_graph}
    \end{subfigure}
       \caption{QAOA benchmark circuit solving the maximum independent set problem for a random Erdos-Renyi graph with $5$ qubits.}
       \label{fig:compute_graph_example}
\end{figure*}
With the overall framework in mind,
this section discusses the first step in circuit cutting -- determine the cutting points.

Figure~\ref{fig:circuit} shows an example QAOA quantum circuit solving the maximum independent set problem for a random Erdos-Renyi graph with $5$ qubits.
Quantum circuits can also be represented as directed acyclic graphs (DAG),
where the vertices are the gates and the edges are the qubit lines connecting the gates.
Hence, locating the cutting points is equivalent to partitioning the quantum gates into certain number of subcircuits.
Note that DAG partition problems only concern with the connection among the vertices,
but the single qubit gates do not affecct such connections.
As a result, the single qubit gates are ignored during cuts finding,
and are simply attributed to the same subcircuit as their closest two-qubit gate neighbor.

Figure~\ref{fig:dag} shows its corresponding DAG representation after removing the single-qubit gates.
The input (green) and output (red) qubit vertices are illustrated for a clearer correspondence with Figure~\ref{fig:circuit}.
However, there is no reason to partition any I/O vertices solely by themselves.
Therefore, the I/O vertices are also safely ignored during cuts finding.

\subsection{MIP Model}
Figure~\ref{fig:cut_dag} shows the DAG graph to partition,
which only includes the two-qubit quantum gates.
We adapt from~\cite{tang2021cutqc} to encode the DAG for our MIP solver.
We denote the $n_C$ subcircuits as $\left\{C_1,\ldots,C_{n_C}\right\}$,
all the qubit lines as edges in $E$,
and all the quantum gates as vertices in $V$.
Furthermore, we define the edge and vertex variables as
\begin{eqnarray*}
    x_{e,c}&\equiv&
    \begin{cases}
        1 & \text{if edge $e$ is cut by subcircuit $c$}\\
        0 & \text{otherwise}
    \end{cases}\\
    y_{v,c}&\equiv&
    \begin{cases}
        1 & \text{if vertex $v$ is in subcircuit $c$}\\
        0 & \text{otherwise}
    \end{cases}\\
    &&\forall{e}\in{E},\forall{v}\in{V},\forall{c}\in{\{1\ldots n_C\}}
\end{eqnarray*}

The numbers of incoming and outgoing cut edges to a subcircuit are modeled as
\begin{eqnarray*}
    I_c&=&\sum_{e\in E}x_{e, c} \times y_{e[1], c}\\
    O_c&=&\sum_{e\in E}x_{e, c} \times y_{e[0], c}
\end{eqnarray*}
The number of quantum gates in each subcircuit is modeled as
\begin{equation*}
    S_c = \sum_{v\in V}y_{v, c}
\end{equation*}

\subsection{MIP Constraints and Objective}
Figure~\ref{fig:compute_graph} shows the subcircuit abstraction obtained after applying some cuts to the DAG.
We call this abstraction the compute graph,
where each subcircuit is a vertex and the edges are the cuts selected.
Each subcircuit is also connected with some output qubit vertices.
We limit the max subcircuit size based on the load factor $\alpha$ and demand that
\begin{equation}
    S_c\leq\alpha|V|,\forall{c}\in{C}\label{MIP:load_constraint}
\end{equation}
In addition, the MIP model contains the corresponding constraints to help define the edge and vertex variables.

Our efficient post-processsing algorithms escape the seemingly exponential cost suggested by Equation~\ref{eq:CutQC},
but they also make the exact post-processing cost of quantum circuit cutting much nuanced and depend on many factors.
We defer the detailed discussions of post-processing the compute graph to Section~\ref{sec:compute_graph_contraction},
specifically its cost in Section~\ref{sec:contraction_cost}.
Let us accept for now that the computation complexity is closely related with the compute graph degree,
which serves as an indirect measure of the computation overhead.
The compute graph degree is just the max number of cuts on any subcircuit,
which is simply the sum of the incoming and outgoing cut edges.
Therefore, our MIP solver seeks to minimize
\begin{equation}
    L\equiv \max_{c\in{\{1\ldots n_C\}}}{\{I_c+O_c\}}\label{MIP:objective}
\end{equation}

Furthermore, the number of subcircuits $n_C$ to partition into cannot be captured as part of the optimization model.
Instead, we run our solver for up to $n$ subcircuits for a $n$-qubit benchmark to obtain several possible solutions.
Results from Section~\ref{sec:compute_graph_contraction} then allow us to predict the exact post-processing cost for each.
The cutting solution with the lowest computation cost is eventually selected.

Overall, the cut search problem is transformed as an integer programming model.
However, this constrained graph partition problem is conjectured to have no efficient solutions in polynomial time~\cite{andreev2006balanced}.
Instead, we utilized the commercial solver Gurobi~\cite{gurobi} to implement our model.
We limit the runtime to $30$ seconds for each candidate $n_C$
for an approximate solution if an optimal solution is not found within the time.
\section{Compute Graph Contraction}\label{sec:compute_graph_contraction}
This section introduces compute graph contraction for an efficient post-processing.
According to Equation~\ref{eq:CutQC},
the final reconstruction of the output of the uncut circuit requires permutating the $\{I,X,Y,Z\}$ indices for the $K$ cut edges,
taking the outer products among the subcircuits, and sum over all the $4^K$ terms.
The naive computation of Equation~\ref{eq:CutQC} quickly becomes the bottleneck for complicated circuits
as it incurs many redundant computations.
This is because subcircuits in general do not connect with all the edges
and hence remain unchanged across many different edge bases.
In addition, it is preferrable to group many outer products as matrix products,
as GPUs usually compute matrix products much faster than the explicit iterations over the rows and columns.
Compute graph contraction hence seeks to eliminate the redundant computations and exploit the more efficient compute kernels.

\subsection{Relation with Tensor Network Contraction}
\begin{figure}[t]
    \centering
    \begin{subfigure}{0.3\textwidth}
        \centering
        \includegraphics[width=\textwidth]{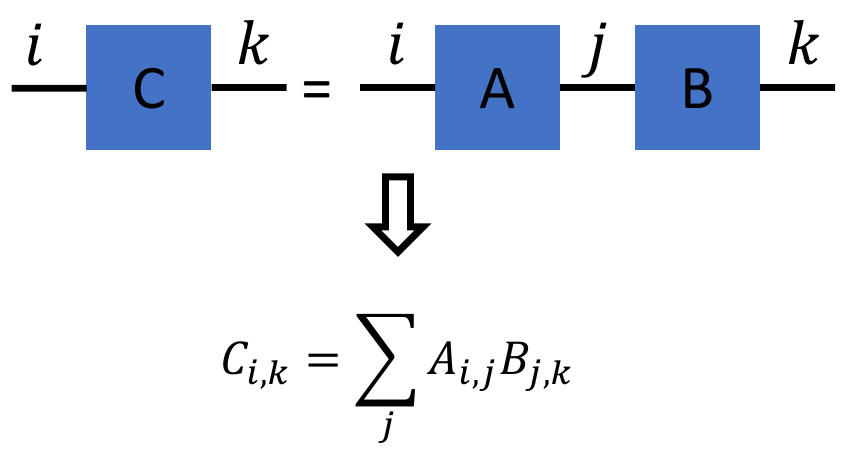}
        \caption{Contracting a pair of tensors means taking the sum over their common index (indices).}
        \label{fig:tensor_contraction}
    \end{subfigure}
    \begin{subfigure}{0.4\textwidth}
        \centering
        \includegraphics[width=\textwidth]{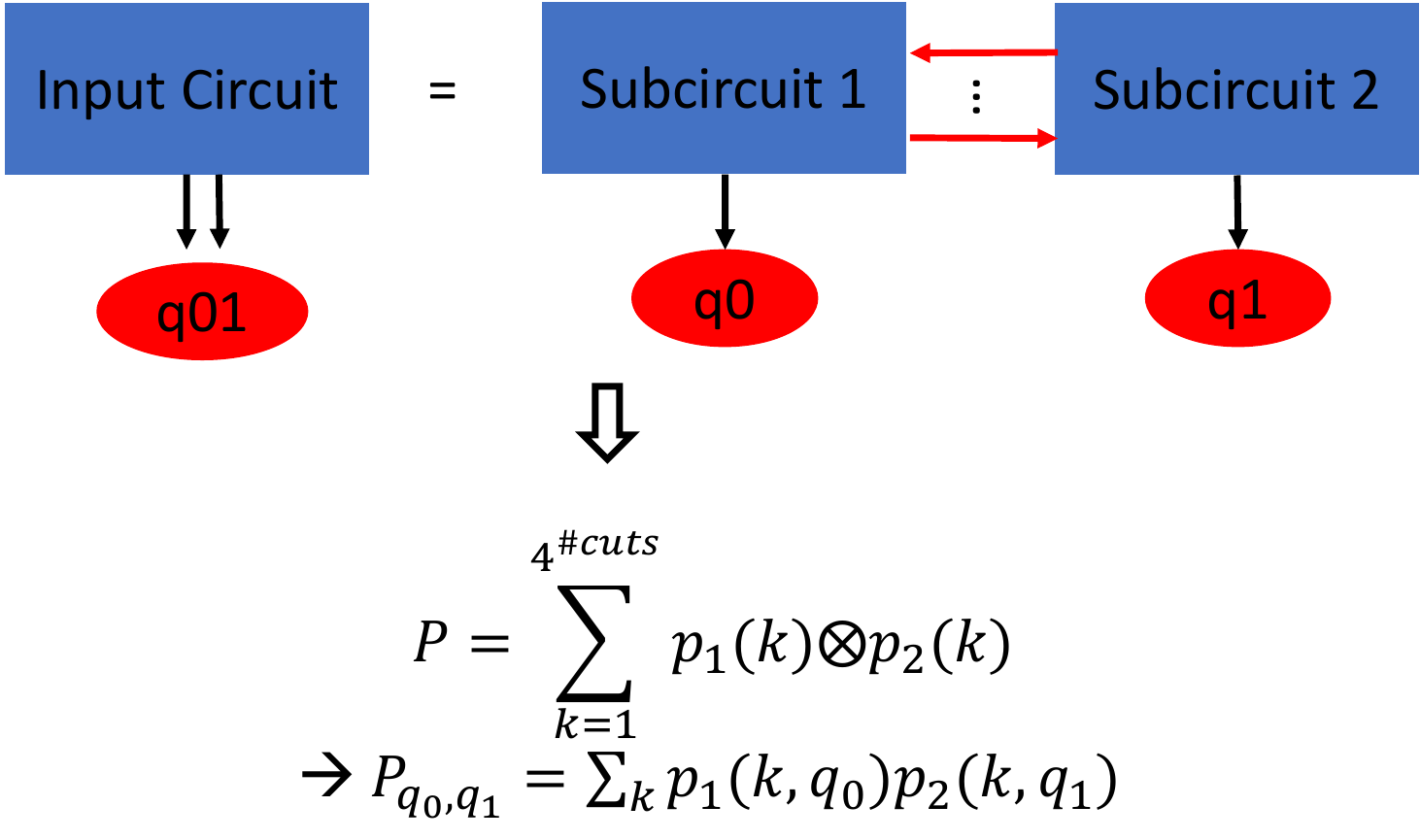}
        \caption{Contracting a pair of subcircuits is equivalent to tensor contraction.
        Each cut edge has a dimension of $4$,
        each output qubit has a dimension of $2$.}
        \label{fig:subcircuit_contraction}
    \end{subfigure}
    \caption{Contracting two subcircuits is equivalent to a pair of tensor contraction.}
    \label{fig:graph_to_tensor}
\end{figure}
\begin{figure}[t]
    \centering
    \begin{subfigure}{0.35\textwidth}
        \centering
        \includegraphics[width=\textwidth]{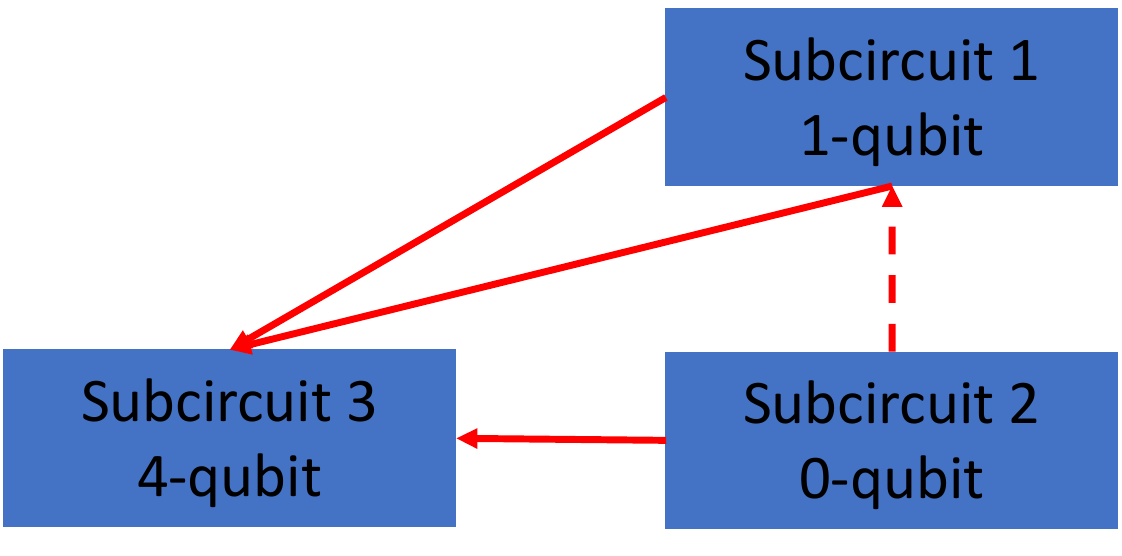}
        \caption{Contracting subcircuits $1,2$.
        Subcircuit $1$ tensor has dimensions of $\{4,4,4,2^1\}$.
        Subcircuit $2$ tensor has dimensions of $\{4,4,2^0\}$.
        Subcircuit $3$ tensor has dimensions of $\{4,4,4,2^4\}$.}
        \label{fig:contraction_step_1}
    \end{subfigure}
    \begin{subfigure}{0.4\textwidth}
        \centering
        \includegraphics[width=\textwidth]{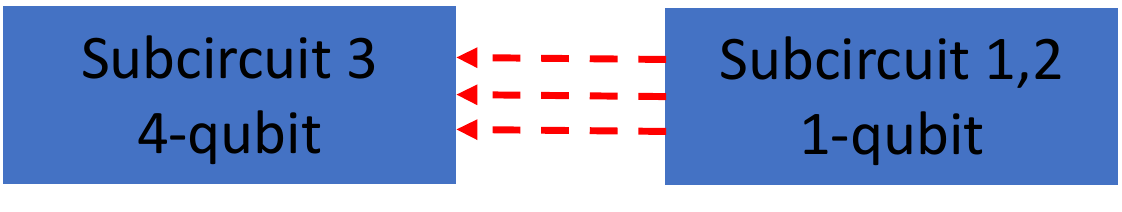}
        \caption{Contracting subcircuit $3$ with the already contracted subcircuits $1,2$.}
        \label{fig:contraction_step_2}
    \end{subfigure}
    \begin{subfigure}{0.4\textwidth}
        \centering
        \includegraphics[width=\textwidth]{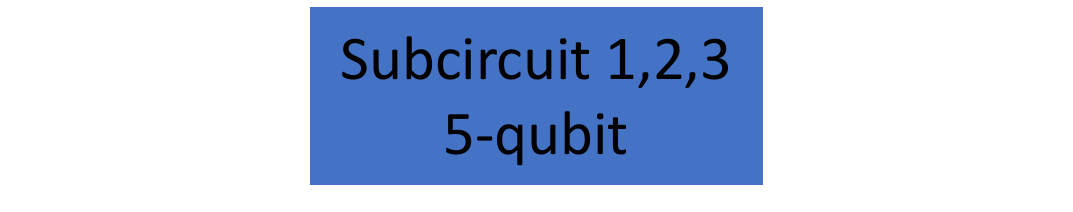}
        \caption{Contraction process finishes.
        The output is a single tensor of dimension $\{2^5\}$,
        as expected from a $5$-qubit input circuit.}
        \label{fig:contraction_step_3}
    \end{subfigure}
       \caption{Contracting the compute graph example from Figure~\ref{fig:compute_graph}.
       The output qubits are abbreviated as the number of qubits in each tensor for brevity.
       The dashed edges are the contraction edges in each step.
       The sub-figures show the best contraction order obtained from Section~\ref{sec:determine_contraction_order}.}
       \label{fig:compute_graph_contraction_example}
\end{figure}
Post-processing the compute graph is in fact equivalent to tensor network contractions,
which have been widely used in classical simulations of quantum systems~\cite{vidal2003efficient,vidal2004efficient,schollwock2011density,verstraete2004matrix}.
Figure~\ref{fig:graph_to_tensor} establishes the equivalence of a pairwise tensor contraction with a compute graph contraction containing two subcircuits.

Figure~\ref{fig:tensor_contraction} shows contracting a pair of tensors,
which are simply multi-dimensional matrices.
It hence represents contracting
a tensor $A$ of shape $\dim(i)$-by-$\dim(j)$ with
a tensor $B$ of shape $\dim(j)$-by-$\dim(k)$
to produce a tensor $C$ of shape $\dim(i)$-by-$\dim(k)$,
where $\dim(\cdot)$ is the dimension of an index.
By examining the definition of the resulting tensor $C$,
it is easy to verify that a pairwise tensor contraction is simply the matrix product $C=AB$.

Figure~\ref{fig:subcircuit_contraction} shows a hypothetical example of contracting a compute graph with two subcircuits.
Each subrcircuit is a tensor,
the cut edges are the common indices to be contracted,
and the output qubits are the output indices.
Each cut edge has a dimension of $4$ because of the $\{I,X,Y,Z\}$ labels to permutate.
Each output qubit has a dimension of $2$ because of the $\ket{0},\ket{1}$ bases.
Explicitly writing out the output indices based off Equation~\ref{eq:CutQC} clearly shows that
the compute graph contraction is exactly a tensor contraction.

Figure~\ref{fig:compute_graph_contraction_example} shows the contraction process for the compute graph in Figure~\ref{fig:compute_graph} as tensor contractions.
Each subcircuit has different number of cuts and number of qubits,
hence different dimensions of tensors at the beginning.
Each step contracts two vertices in the compute graph.
The contraction process finishes when all the vertices are contracted into a single vertex.
The final output of compute graph contraction is a single tensor
of dimension $\{2^n\}$ for a $n$-qubit circuit when using full state reconstruction.

Furthermore, when we apply states merging to large circuits beyond classical memory,
the beginning tensors have their last dimensions as however many bins each subcircuit has.
The final output of compute graph contraction is instead a single tensor of dimension $\{M\}$.

\subsection{Memory Requirement and Compute Cost}\label{sec:contraction_cost}
The memory requirement of the post-processing comes from two parts.
First, we need to store the input subcircuit tensors.
The tensor of subcircuit $i$ has $4^{\#cuts_i}\times 2^{\#qubits}$ ($4^{\#cuts_i}\times\#bins_i$) floats
when employing full state (states merging).
Second, we need to store all the intermediate tensor products during the contraction.

The compute cost of the post-processing can be captured by the floating point multiplications required in all the contraction steps.
The number of multiplications of a pair of contraction is simply the product of all the dimensions involved.
For example, Figure~\ref{fig:tensor_contraction} requires $\dim(i)\times\dim(j)\times\dim(k)$ multiplications.
Similarly, Figure~\ref{fig:subcircuit_contraction} requires $2^1\times 4^{\#cuts}\times 2^1$ multiplications.

Now we can easily predict the memory and compute overhead of a compute graph without actually performing the computations.
The input subcircuit tensors for the example contraction step in Figrue~\ref{fig:contraction_step_1} require total
$4^3\times 2^1 + 4^2\times 2^0 + 4^3\times 2^4 = 1168$ float numbers
and $(4^2\times2^1)\times(4^1)\times(4^1\times2^0)=512$ multiplications.
The contraction step in Figure~\ref{fig:contraction_step_2} requires $4^3\times2^4+4^3\times2^1=1152$ float numbers
and $2^4\times4^3\times2^1=2048$ multiplications.

In contrast, the naive computation of Figure~\ref{fig:contraction_step_1} based on Equation~\ref{eq:CutQC} requires
$4^4\times(2^1\times2^0+2^1\times2^4)=8704$ multiplications, a $3.4\times$ extra overhead.

It is now clear that both the memory and compute cost of compute graph contraction are positively correlated with the number of cuts on each subcircuit during every contraction step.
This explains our MIP solver objective choice in Equation~\ref{MIP:objective}.
In order to keep the post-processing runtimes reasonable,
we limit the compute graph degree to be $15$.

\subsection{Two Level Index Slicing}
\begin{figure}[t]
    \centering
    \includegraphics[width=\linewidth]{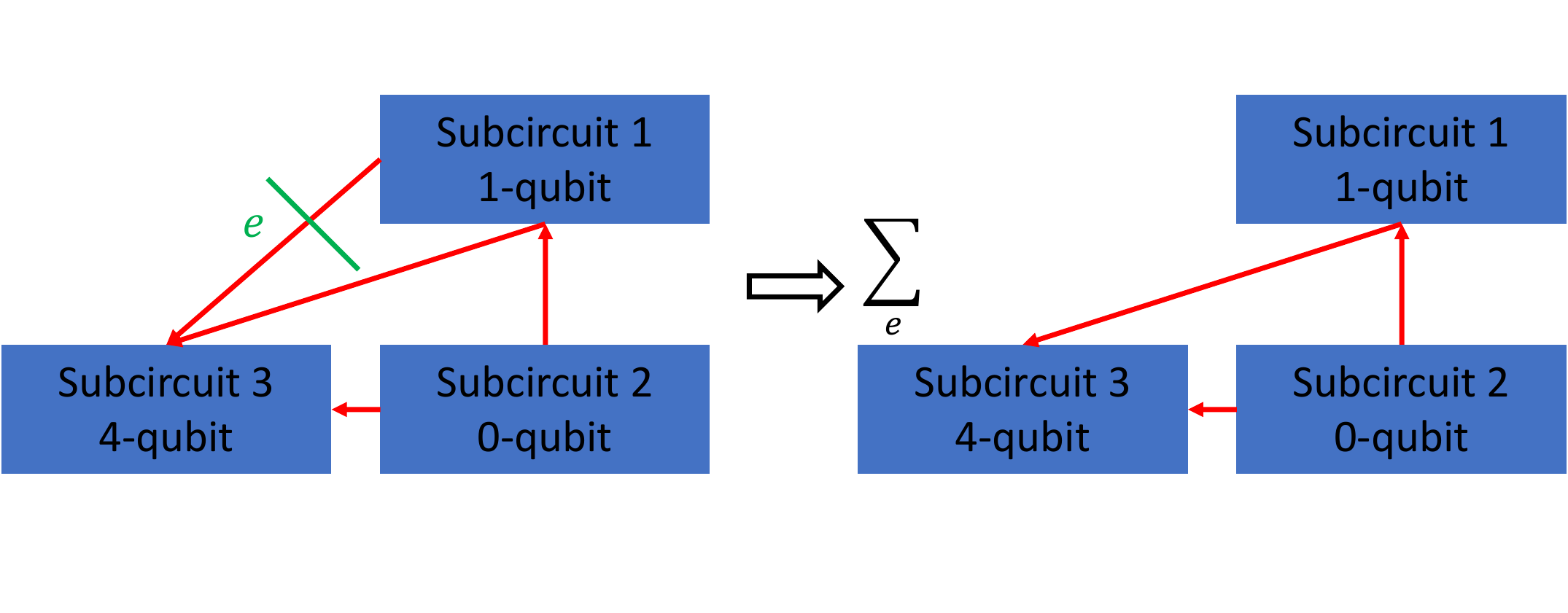}
    \caption{Slicing one cut edge to express the compute graph as a summation of smaller compute graphs.}
    \label{fig:slicing}
\end{figure}
It is possible that complicated compute graphs with many cuts require input subcircuit tensors too large to fit in the memory.
Hence we implement an index slicing strategy prior to their contraction to reduce the input tensor sizes.
Index slicing is to explicitly write out certain indices and express the full tensor network as a summation of smaller tensor networks.
In the context of compute graph, it is to explicitly write out certain cut edges.
For example, slicing one edge in Figure~\ref{fig:slicing} reduces the input subcircuit tensor sizes to
$4^2\times2^1+4^2\times2^0+4^2\times2^4=304$ float numbers but produces $4$ smaller compute graphs to contract.
Our heuristics keep slicing the index that reduces the overall tensor sizes the most,
until the overall sizes fit in the memory specified.
The heuristics hence produce a set of smaller compute graphs that we explicitly iterate over.

In addition, the intermediate tensor products during contractions may also exceed memory.
We further applied the index slicing algorithm from the CoTenGra software~\cite{gray2021hyper} to each sub compute graph generated from the first level of slicing.

\subsection{Determine Contraction Order}\label{sec:determine_contraction_order}
In tensor network contractions, different contraction orders significantly affect the overhead,
sometimes even introduce orders of magnitude differences.
In circuit cutting,
the contraction order of the subcircuits only affects the reconstruction up to a permutation of the qubit order,
and has no effect on the quantum state probability accuracy.
However, finding the optimal contraction order for general compute graphs is hard.
In fact, it is an active area of research and of broader interest across many disciplines~\cite{robertson1991graph,markov2008simulating,orus2014practical}.
We used the CoTenGra software~\cite{gray2021hyper} to compile the contraction order for each sub compute graph.
\section{Methodology}
This section introduces the various backends, benchmarks, and metrics.

\subsection{Backends}\label{sec:backends}
The following three backends are involved in the experiments:
\begin{enumerate}
    \item Quantum: Direct evaluation of quantum circuits on a powerful enough QPU without circuit cutting;
    \item Cut: The standard ScaleQC framework in full state;\label{backend:cut}
    \item Cut\_M: Cut with states merging.
    This is equivalent to full state evaluation when states merging runs to completion.
    We run the max number of $mn/\log_2{M}$ recursions to guarantee completion.\label{backend:cut_m}
\end{enumerate}

Large circuits tend to have larger subcircuits as well,
but the Noisy Intermediate Scale Quantum (NISQ) devices nowadays are still too small and noisy for any meaningful experiments at even medium sizes.
Hence we use random numbers as the subcircuit output.
Although it does not produce any useful circuit outputs,
it does not affect the post-processing runtime results of the experiments.
We expect more reliable QPUs to enable a full implementation of the ScaleQC framework.

We use a single compute node with $64$ CPUs equipped with a single Nvidia A$100$ GPU
for all the cuts finding and classical post-processing.
\subsection{Benchmarks}\label{sec:benchmarks}
We used the following circuits as benchmarks:
\begin{enumerate}
    \item $BV$: Bernstein-Vazirani circuit~\cite{bernstein1997quantum} solves the hidden substring problem.
    Has $1$ solution state.
    \item $Regular$: Quantum Approximate Optimization Algorithm solves the maximum independent set problem for random $3$-regular graphs~\cite{saleem2020approaches}.
    With the proper hyperparameters, this circuit produces $1$ solution state.
    However, the full QAOA training process to solve the hyperparameters is beyond the scope of this paper.
    We use random hyperparameters with the same circuit structure.
    \item $Erdos$: The same algorithm as $Regular$ but for random Erdos-Renyi graphs.
    \item $Supremacy$: Random quantum circuits adapted from~\cite{boixo2018characterizing}.
    Deeper versions of the circuit was used by Google to demonstrate quantum advantage~\cite{arute2019quantum}.
    Our paper uses depth $(1+8+1)$ at various sizes.
    As it does not have well-defined solution states,
    we run $mn/\log_2{M}$ recursions to produce more samples.
    \item $AQFT$: Approximate Quantum Fourier Transform~\cite{barenco1996approximate} that is expected to outperform the standard QFT circuit under noise.
\end{enumerate}
All of our benchmarks are examples of circuits and routines that are expected to demonstrate quantum advantage over classical computing on the Quantum backend.
ScaleQC aims to demonstrate the middle ground where the overall runtime is slower than Quantum but still beyond the classical reach,
while the quantum and classical computing resources requirements are much lower.
\subsection{Metrics}
There are two key metrics this paper looks at, namely runtime and quantum resources.

\textbf{Runtime, faster is better}:
For Quantum, it is the end-to-end runtime on a standalone QPU, which is the best runtime possible.
The Cut and Cut\_M backends are slower.

The ScaleQC runtime is the end-to-end runtime except time spent on QPUs in Algorithm~\ref{alg:framework}.
The NISQ QPUs nowadays are small, slow and too noisy for any practical purposes.
As we expect the practical ScaleQC applications to be used with medium sized reliable QPUs in the near future,
it is irrelevant to profile the NISQ QPU runtime now.
Furthermore, multiple small QPUs can be used in parallel to reduce the runtime.
In addition, the runtime advantage of QPUs over CPUs will be even more significant for larger circuits.
We expect the framework to offer more significant advantages over classical methods as larger and more reliable QPUs become available.

\textbf{Quantum Area, smaller is better}:
The quantum resources requirement is loosely defined as the product of the circuit width and depth,
called the `quantum area'.
Classical simulations require $0$ quantum area as it does not use QPUs at all.
For Quantum, it is simply the product of the number of qubits and the circuit depth of the input quantum circuit.
For the various ScaleQC backends, it is defined for the largest subcircuit produced from cutting.
The rationale is that QPUs must be able to support the workloads with a certain quantum area at high accuracy to produce accurate results.
While many hardware and software factors affect the QPUs' ability to support quantum workloads,
a smaller quantum area generally puts less burden on the quantum resources.
\section{Experiment Results}\label{sec:results}
\subsection{Runtime}
\begin{figure}[t]
    \centering
    \includegraphics[width=\linewidth]{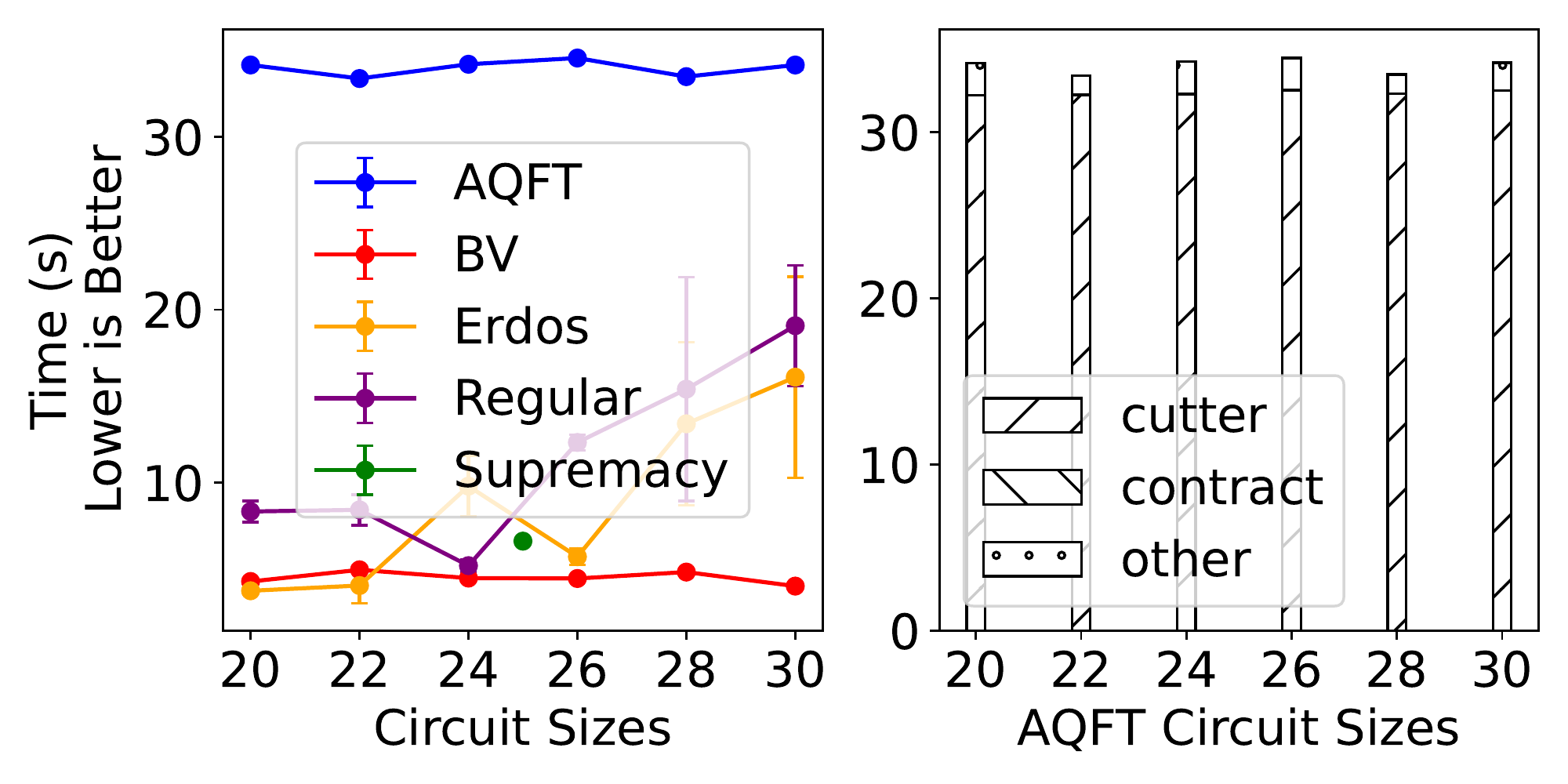}
    \caption{Full state runtimes.
    Max load $\alpha=0.4$.
    Each data point is the average of $5$ trials.
    (Left Panel) The runtimes of $20$-$30$ qubits benchmarks.
    The error bars represent the standard deviations.
    (Right Panel) The runtime breakdown for the $AQFT$ benchmark.
    The majority of the runtime is from the cut searching.}
    \label{fig:full_state_runtime}
\end{figure}

When the size of the Hilbert space of the quantum circuits is still within the memory limit,
it is possible to perform full state evaluation.
This cutoff depends on the particular classical backend available to the users.
We use $30$ qubits as the cutoff in this paper, which is roughly $10^9$ states.
However, since each qubit doubles the memory requirement,
there is little value in pushing this cutoff point to the extreme.

Figure~\ref{fig:full_state_runtime} plots the benchmark circuits $20\to30$ qubits,
subject to a max load of $\alpha=0.4$.
We observe that the majority of the runtime is spent on searching the cuts.
For example, the runtime breakdown of the $AQFT$ benchmarks shows that nearly all of the postprocessing runtime comes from the cut searching.
This pre-processing overhead can be significant for medium sized benchmarks due to their short overall runtime.
As a result, it might not be worth the overhead to use circuit cutting for medium circuits.
Admittedly, classical simulators operate relatively fast and are usually adequate for such small benchmarks.
As a reference, the widely used Qiskit~\cite{Qiskit} simulator takes about $30\to50$ seconds to run the $30$-qubit benchmarks on the same classical backend.
ScaleQC is able to produce comparable runtime even for the small benchmarks where classical simulation excels.

While finding the optimal cut solution can be difficult for large circuits,
our experiments capped the solver runtime to produce high quality solutions.
The cut searching step can hence be viewed as a nearly constant pre-processsing overhead for large benchmarks.
In fact, ScaleQC has more significant runtime advantages for larger circuits,
where the cut searching overhead no longer bottlenecks the framework.
Furthermore, ScaleQC with states merging becomes necessary to deal with the exponentially increasing state space.

\begin{figure}[t]
    \centering
    \includegraphics[width=.7\linewidth]{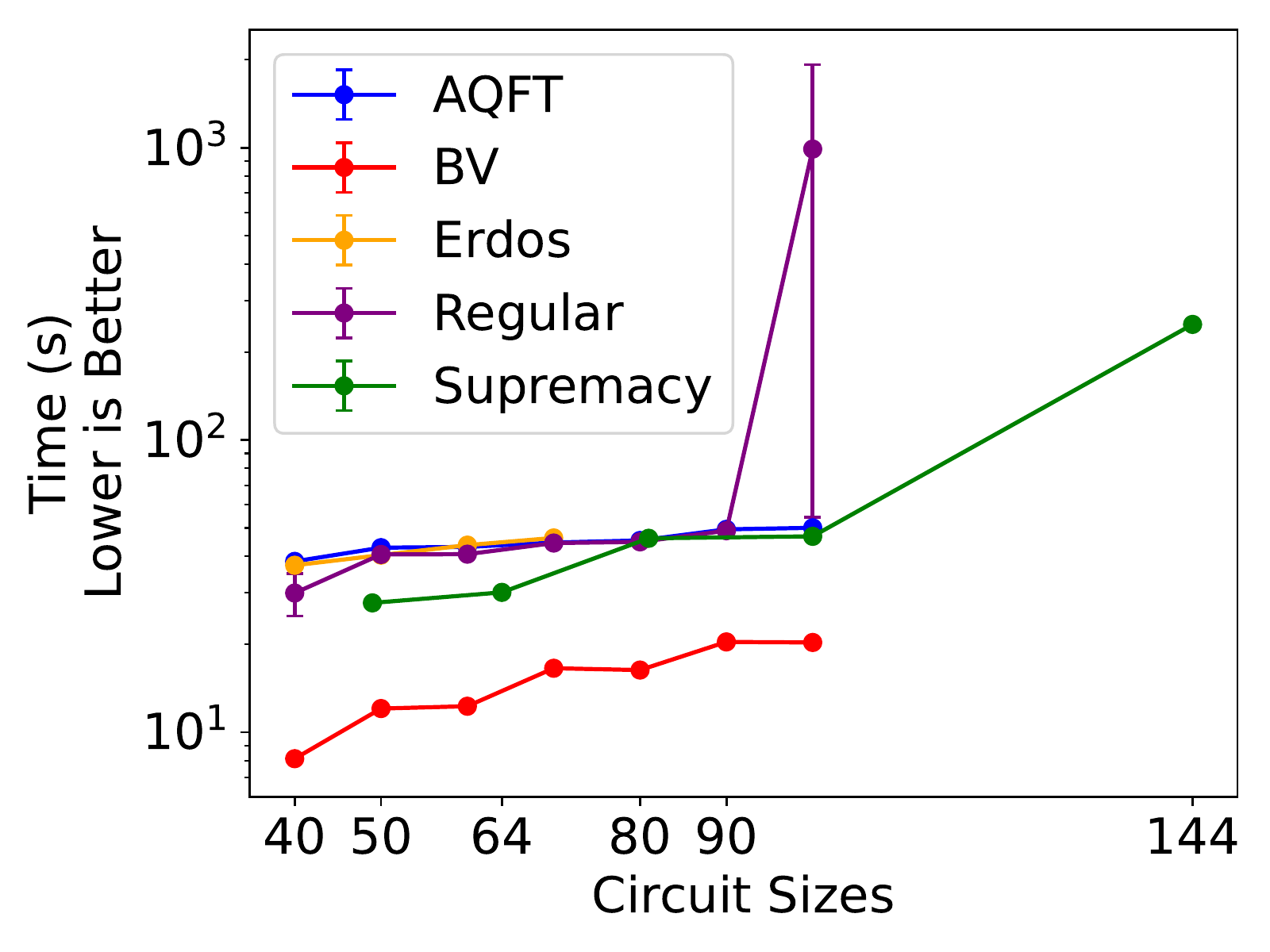}
    \caption{States merging for large circuits.
    Max load $\alpha=0.8$ for $Regular$ and $Erdos$, $\alpha=0.5$ for the rest.
    Max bins $M=2^{20}$.
    Each data point is the average of up to $5$ trials.
    Cut\_M runs $n/\log_2{M}$ recursions (i.e. $5$ for the $100$-qubit benchmarks) to find at least one solution.
    Cut\_M scales well and is able to run circuits significantly beyond the classical simulation limit.}
    \label{fig:states_merging_runtime}
\end{figure}

Figure~\ref{fig:states_merging_runtime} plots the runtime scalability for large benchmarks,
subject to a max load of $\alpha=0.8$ for the two QAOA benchmarks and $\alpha=0.5$ for the rest.
We set $M=2^{20}$ bins.
Each experiment runs $n/\log_2{M}$ recursions to find at least one solution.
The runtime will be the same to reconstruct $Mn/\log_2{M}$ states for the $Supremacy$ benchmark,
which is about $7$ million states for the $144$($12\times 12$)-qubit circuit.
Instead, states merging obtains more information about the entire state space under the same time.

\begin{figure}[t]
    \centering
    \includegraphics[width=.7\linewidth]{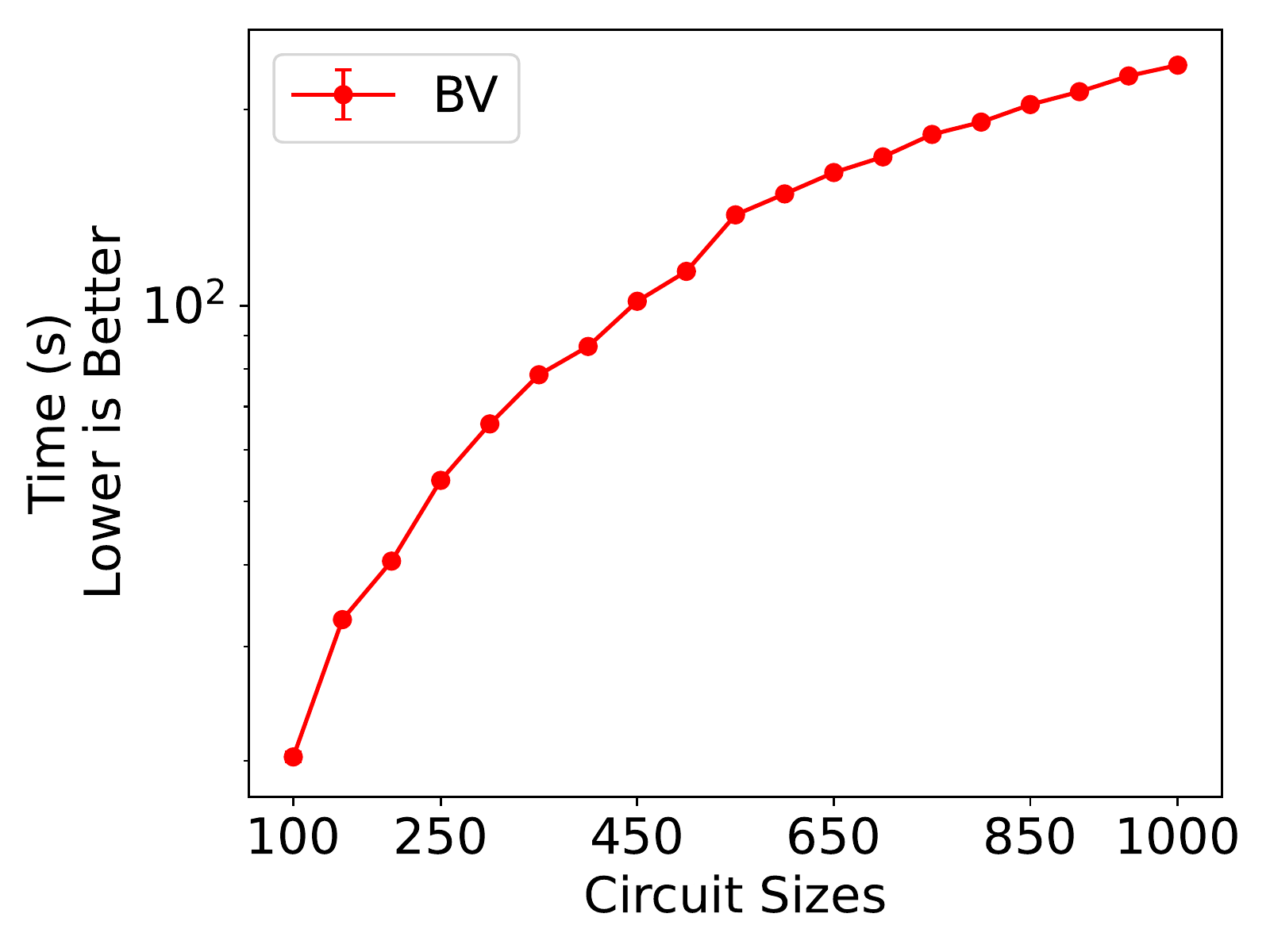}
    \caption{States merging for large BV circuits.
    Max load $\alpha=0.5$.
    Max bins $M=2^{20}$.
    Each data point is the average of $5$ trials.
    Cut\_M runs $n/\log_2{M}$ recursions (i.e. $50$ for the $1000$-qubit benchmark) to find the unique solution.}
    \label{fig:states_merging_runtime_bv}
\end{figure}

Figure~\ref{fig:states_merging_runtime_bv} plots the runtime scalability for the BV benchmark up to $1000$ qubits,
subject to a max load of $\alpha=0.5$ and $M=2^{20}$ bins.
Each experiment runs to completion to find the unique hidden string solution state.

The various benchmarks scale well as circuits get larger.
The $Erdos$ benchmark appears to be the hardest and no solutions were found above $80$ qubits for the maximum of $15$ compute graph degree.

\subsection{Resources Estimations}
\begin{figure}[t]
    \centering
    \includegraphics[width=.7\linewidth]{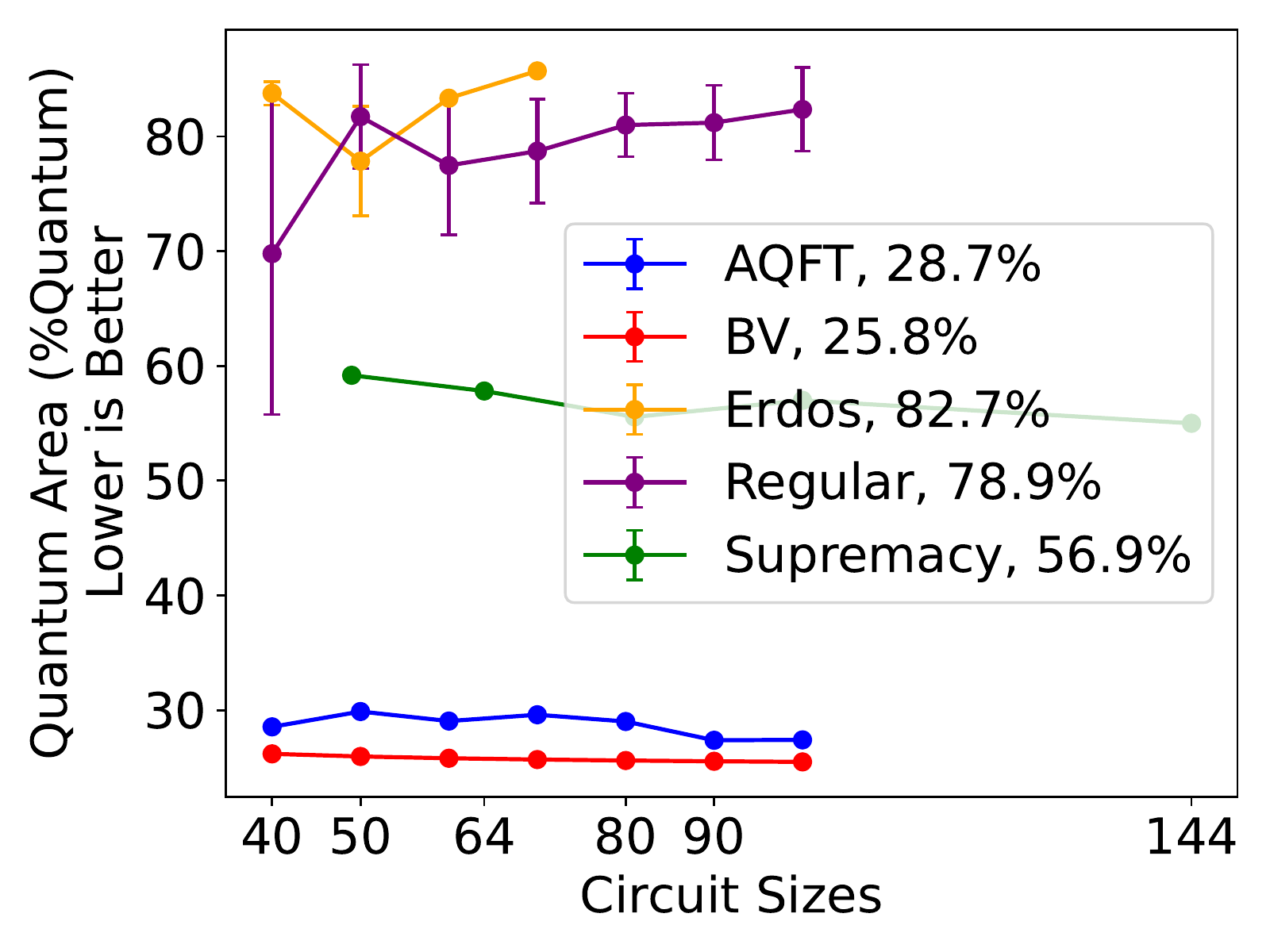}
    \caption{Quantum areas comparisons for the same experiments in Figure~\ref{fig:states_merging_runtime}.
    The numbers in the legends indicate the average quantum area of Cut\_M as a percentage over Quantum.
    Cut\_M requires QPUs to support at most $83\%$ of the quantum area at the expense of the classical runtimes in Figure~\ref{fig:states_merging_runtime}.}
    \label{fig:quantum_area}
\end{figure}

\begin{figure}[t]
    \centering
    \includegraphics[width=.7\linewidth]{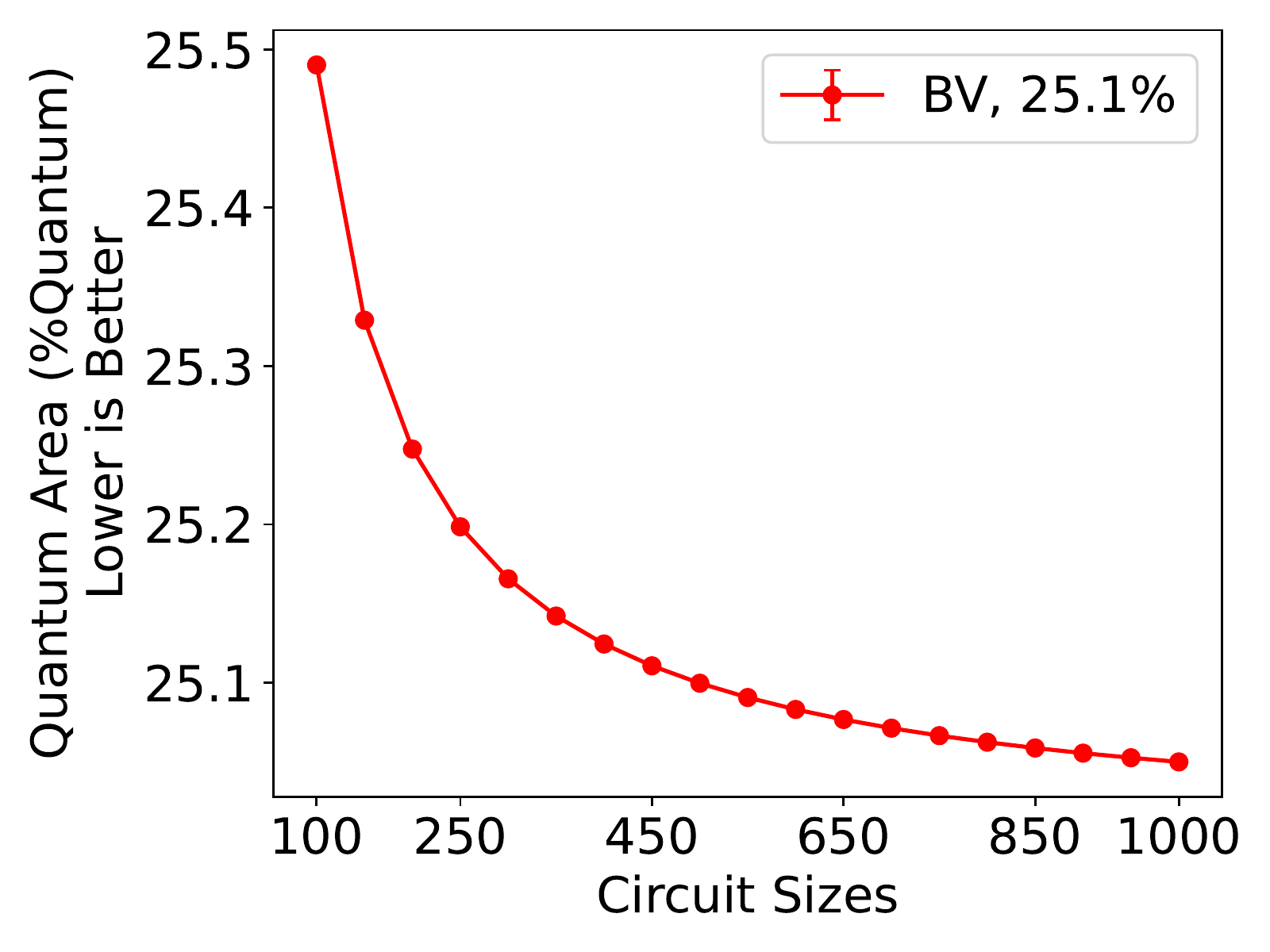}
    \caption{Quantum areas comparisons for the same experiments in Figure~\ref{fig:states_merging_runtime_bv}.
    Cut\_M requires QPUs to support about $25\%$ of the quantum area at the expense of the runtimes in Figure~\ref{fig:states_merging_runtime_bv}.}
    \label{fig:quantum_area_bv}
\end{figure}

ScaleQC also requires much less powerful QPUs than Quantum since it just needs to support the smaller subcircuits.
Figure~\ref{fig:quantum_area} plots the quantum areas of Cut\_M as the percentage of Quantum for the same experiments in Figure~\ref{fig:states_merging_runtime}.
Figure~\ref{fig:quantum_area_bv} similarly plots the quantum areas of the large BV benchmarks in Figure~\ref{fig:states_merging_runtime_bv}.
Quantum simply requires the area of the input benchmark circuit.
Cut\_M requires QPUs to support the quantum area of the largest subcircuit from cutting.
With $\alpha=0.5$($0.8$), Cut\_M requires QPUs to support at most about $57\%$($83\%$) of the quantum area to run the various benchmarks,
at the expense of the postprocessing runtimes in Figures~\ref{fig:states_merging_runtime}~and~\ref{fig:states_merging_runtime_bv}.

Lower quantum area requirements translate to the ability to tolerate smaller and noisier NISQ devices.
Furthermore, in the fault tolerant regime,
lower quantum areas translate to a looser requirement on the logical qubit error rate.
Depending on different quantum error correction solutions,
this implies much reduced physical qubit counts and error threshold requirements.
In addition, compiling and decoding the smaller subcircuits are also much easier than larger circuits,
which put much less burden on the quantum software development~\cite{tan2020optimality}.
\subsection{Compare Against Classical Simulation}
\begin{figure}[t]
    \centering
    \includegraphics[width=\linewidth]{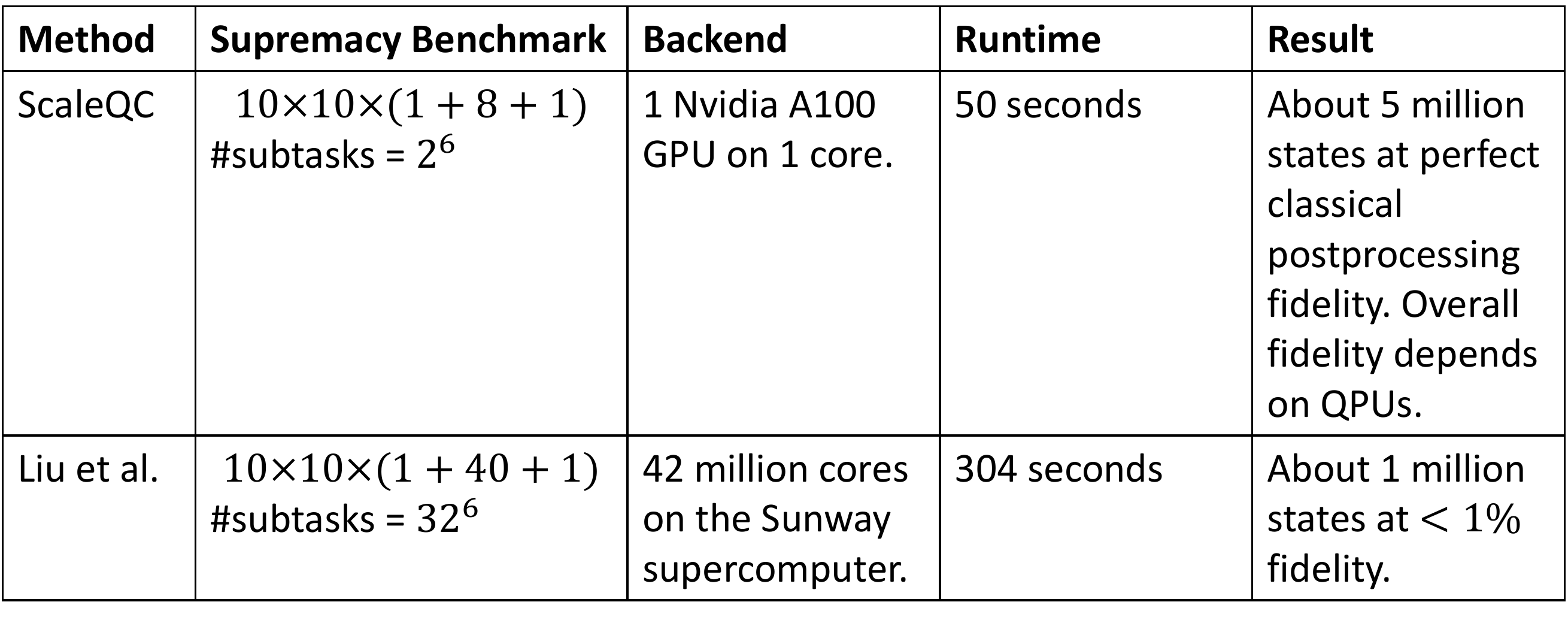}
    \caption{Compare against~\cite{liu2021closing}.
    We estimate~\cite{liu2021closing} requires $42\times10^6\times\frac{2^6}{32^6}\times100\times5\approx 1250$ cores and $300$ seconds for our benchmark.
    Each recursion of ScaleQC produces $1$ million more states at minimal extra runtime cost.
    However,~\cite{liu2021closing} requires the same time for each $1$ million more states.}
    \label{fig:comparison}
\end{figure}

\begin{figure}[t]
    \centering
    \includegraphics[width=.8\linewidth]{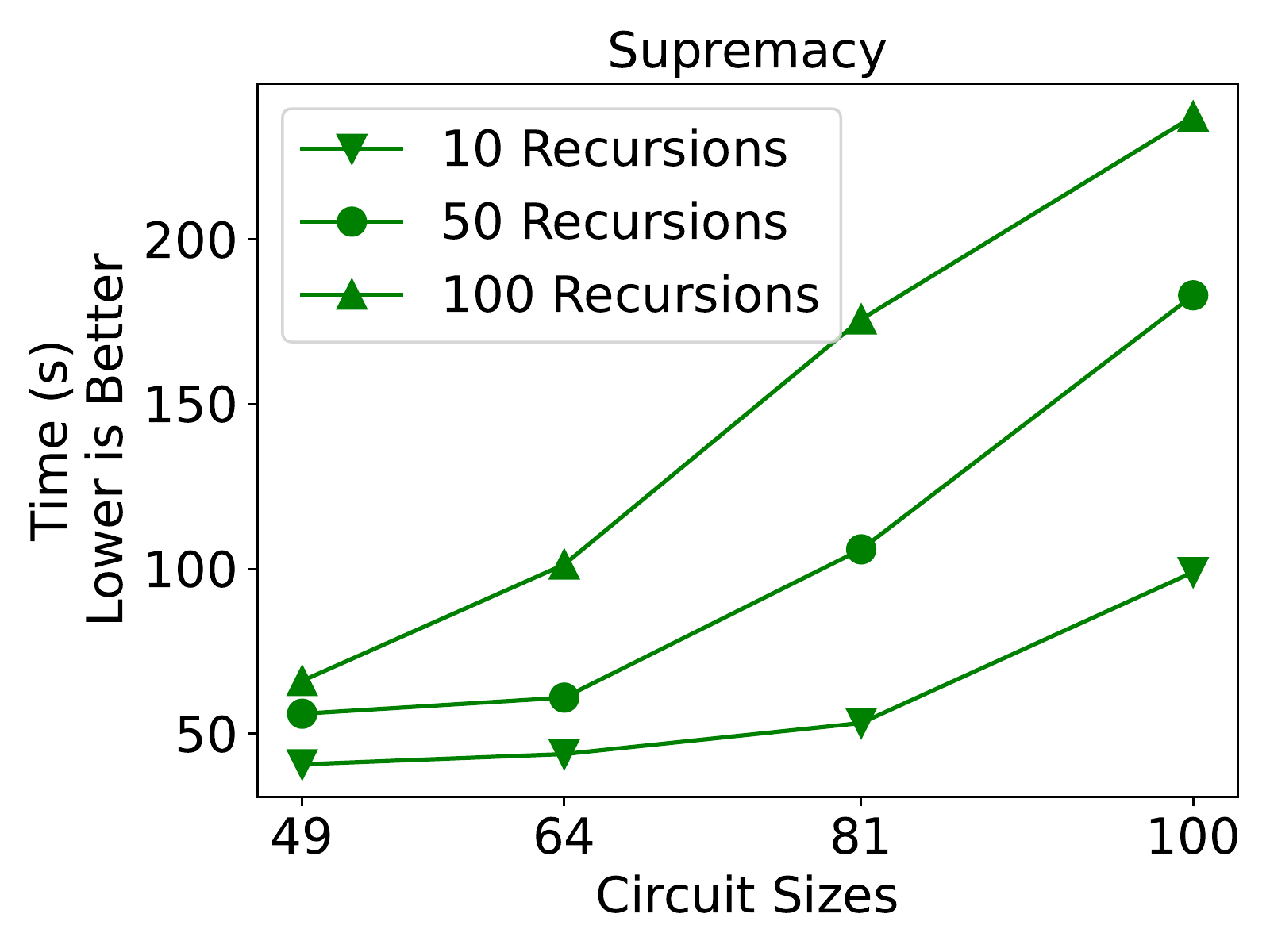}
    \caption{Running the $Supremacy$ benchmark for varying number of recursions.
    Max load $\alpha=0.5$, max bins $M=2^{20}$.
    Since each states merging recursion is fast,
    adding more recursions do not increase the runtime too much.}
    \label{fig:recursion_runtime}
\end{figure}

There are lots of purely classical simulation works~\cite{wu2019full,chen201864,haner20175,pednault2019leveraging}.
The state-of-the-art tensor network based method~\cite{liu2021closing} slices a deeper $100$-qubit $Supremacy$ circuit into $32^6$ ``subtasks''
to fit onto almost $42$ million cores on the Sunway supercomputer.
As a comparison, our shallower benchmark will be sliced into $2^6$ such ``subtasks'' by the same metric.

Figure~\ref{fig:comparison} shows a qualitative comparison for the most related benchmark circuit.
To maintain their $300$ seconds runtime, we estimate~\cite{liu2021closing} to require $42\times10^6\times\frac{2^6}{32^6}\approx2.5$ cores for $1$ million samples at $<1\%$ fidelity,
or $2.5\times5\times100=1250$ cores for $5$ million samples at the same perfect postprocessing fidelity as ScaleQC shows in Figure~\ref{fig:states_merging_runtime}.

We can easily modify the states merging framework to instead produce $1$ million arbitrary states in each recursion.
Figure~\ref{fig:recursion_runtime} runs various $Supremacy$ benchmarks for $10,50,100$ recursions.
Specifically, $5$ recursions of the $100$-qubit $Supremacy$ benchmark takes about $50$ seconds in Figure~\ref{fig:states_merging_runtime}.
Meanwhile, $100$ recursions of the same benchmark only takes about $160$ seconds in Figure~\ref{fig:recursion_runtime}.
The little runtime increase for more recursions shows that the majority of the runtime is the one-time initial cut searching.
Furthermore, the cut searching allows our method to generalize to any benchmarks.

In contrast,~\cite{liu2021closing} manually pre-determines the slicing strategy specifically but only for the $Supremacy$ benchmark.
It also needs to spend the same runtime for every $1$ million more samples calculated.

Deeper and more complicated benchmarks are going to require more than a single GPU.
ScaleQC relies on the continued developments of HPC techniques to port to scalable parallel computing backends.
\section{Related Work}\label{sec:related_work}
Many quantum compiler works exist to improve the performance of standalone QPUs to evaluate quantum circuits~\cite{nannicini2021optimal,tan2020optimality,murali2019noise,ding2020systematic,murali2020software}.
Quantum error correction is the key to build reliable QPUs~\cite{fowler2012surface,bravyi2012subsystem,javadi2017optimized,yoder2017surface,litinski2019game}.
QAOA uses classical computing to tune quantum circuit hyper-parameters to solve optimization problems~\cite{farhi2014quantum,tomesh2021coreset}.
However, they still rely entirely on QPUs to compute the quantum circuits.

Prior circuit cutting implementations~\cite{tang2021cutqc,tang2022cutting} rely on parallelization techniques for faster compute
while performing direct reconstruction of Equation~\ref{eq:CutQC} with high overhead.
The various post-processing algorithms proposed in this paper go beyond what is possible from such techniques
and reduce the overhead itself.
\section{Conclusion}
This paper overcomes the classical runtime and memory scalability challenges for hybrid computation
via novel post-processing algorithms and develops the corresponding cuts searching algorithm.
By distributing large quantum workloads to quantum and classical processors,
we demonstrate up to $1000$-qubit quantum circuits running on both QPUs and GPUs,
which is significantly beyond the reach of either platform alone and previous hybrid workflows.
As ScaleQC bridges the classical and the quantum technologies and paves the way for hybrid computations,
its future developments naturally benefit from advancements on both sides.
\section*{Acknowledgements}
We thank Johnnie Gray, Stojche Nakov, Naorin Hossain, Fran\c{c}ois Pellegrini and George Bosilca for helpful discussions and feedback.

Funding acknowledgements:
This work is partly funded by EPiQC, an NSF Expedition in Computing, under grants CCF-1730082/1730449.
This work is partly based upon work supported by the U.S. Department of Energy, Office of Science, National Quantum Information Science Research Centers, Co-design Center for Quantum Advantage (C2QA) under contract number DE-SC0012704.
This material is based upon work supported by (while Martonosi was serving at) the National Science Foundation.

\bibliographystyle{plain}
\bibliography{refs}
\end{document}